\documentclass[a4paper,11pt,oneside,english]{article}

\usepackage[english]{babel}

\usepackage[centertags]{amsmath}
\usepackage{amsfonts}
\usepackage{amssymb}
\usepackage{color}
\usepackage{amsthm}
\usepackage{newlfont}
\usepackage[applemac]{inputenc}
\usepackage[dvips]{graphicx}

\begin{document}
\title{Minimal Agent Based Model for Financial Markets II: Statistical Properties of the Linear and Multiplicative Dynamics}
\author{V. Alfi$^{1,2}$, M. Cristelli$^1$, L.Pietronero$^{1,3}$, A. Zaccaria$^1$\\\\
$^1$ \small Universit\`a ``La Sapienza'', P.le A. Moro 2, 00185, Roma, Italy\\
$^2$ \small Centro ``E. Fermi'', Compendio Viminale, 00184, Roma, Italy\\
$^3$ \small  ISC-CNR, V. dei Taurini 19, 00185, Roma, Italy
}
\maketitle

\abstract{We present a detailed study of the statistical properties of the Agent Based Model introduced in paper I \cite{paperoI} and of its generalization to the multiplicative dynamics. The aim of the model is to consider the minimal elements for the understanding of the origin of the Stylised Facts and their Self-Organization. The key elements are fundamentalist agents, chartist agents, herding dynamics and price behavior. The first two elements correspond to the competition between stability and instability tendencies in the market. The herding behavior governs the possibility of the agents to change strategy and it is a crucial element of this class of models. The linear approximation permits a simple intepretation of the model dynamics and, for many properties, it is possible to derive analitical results. The generalized non linear dynamics results to be extremely more sensible to the parameter space and much more difficult to analyse and control. The main results for the nature and Self-Organization of the Stylised Facts are, however, very similar in the two cases. The main peculiarity of the non linear dynamics is an enhancement of the fluctuations and a more marked evidence of the Stylised Facts. We will also discuss some modifications of the model to introduce more realistic elements with respect to the real markets.}
\section{Introduction}
In the preceding paper (paper I \cite{paperoI}) we have introduced a minimal agent based model to discuss the origin and self-organization of the Stylised Facts (SF), which are a common characteristics of all price time series. The model is based on four essential elements (first introduced by Lux and Marchesi (LM) \cite{Lux:1999,LM}): fondamentalists (F), chartists (C), herding, price behavior. 
Our aim, in the construction of the agent based model, has been to look for the maximun simplification and reduction of free parameters in order to achieve a detailed understanding of the dynamics of the system. This is achieved by a series of simplifications, discussed in paper I \cite{paperoI} which, however, maintain the essential elements of the model. \\
The construction of an agent-based model represents necessarily a compromise between simplicity and realism, many of the models which have been proposed, are able indeed to reproduce some of the SF. However, often the origin of these properties remains hidden in the many parameters of the models. In this perspective the addition of too many elements ,which make the model more realistic, may in the end also represent an element of confusion for a detailed understanding of the dynamics of the model. This point of view has been argued very clearly in three recent reviews \cite{reviu, hommes, lebaron}. These authors in fact point out that it would be highly desirable to introduce a model which is as simple as possible but it is still to able to repoduce the SF. This is precisely the point of view we have adopted in this and in previous paper where we describe in detail this minimal model. Once the essential elements are clarified it is certainly possible to generalise the model by introducing more realistic elements, and this will be the object of our future work.  \\
In paper I \cite{paperoI} we have introduced and discussed the essential features of the model. The SF are shown to correspond to finite size effect, with important conceptual and pratical implications. We propose that the self-organised state is linked to a threshold in the agent activities which strongly modulates the number of active agents. This leads to a feedback mechanism which triggers spontaneously the system towards the self-organized intermittency state. This state is not really critical, in the sense of statistical physics, because it is related to finite size effects which can, however , occur at different scales.\\
In the present paper we are going to examine in details the statistical properties of our model in various directions. We will try, as much as possible, to derive analitical results. This possibility is much easier if one considers a linear dynamics for the price instead of the more realistic multiplicative dynamics. The two cases essentially coincide for small price variations and we will consider in details their relation and possible discrepancies. \\
The paper is organized as it follows.\\
Section 2 provides some analytical results in two important regimes of the model and in section 3 we investigate the non trivial diffusional properties. In section 4 we focus on the behavior of the tails of the return probability density function (pdf) when the time lag at which the calculation of returns is performed becomes larger and larger. The results of Section 4 suggest a simple approximation that explains how the tails of return pdf are generated by the model. The approximation also allows us to test for which SF property the population dynamics plays a crucial role and this will be discussed in Section 5. In section 6 we introduce an additional RW performed by the fundamental price and this will lead to interesting crossover phenomena.  In Section 7 we consider the multiplicative version of this minimal ABM, and we discuss some non trivial aspects with respect to the linear dynamics. We conlude in section 8 with a short summary of the results of our analysis also with respect to the empirical results, with a discussion of future developments of the model. In appendix A we discuss the implication of the use of diverse definition of the returns. In appendix B we discuss the statistical significance of dimension of a dataset with respect to various properties. Finally in appendix C we discuss data collapse for the return pdf tails and this property can only occur over limited interval of parameters.

\section{Chartists or Fundamentalists}
In paper I \cite{paperoI} and in \cite{paperNP} we have highlighted the key role played by the coexistence of two kind of agents (i.e. two kind of market strategies) and by the transitions from chartists to fundamentalists and viceversa. Anyway our model also shows some interesting aspects in the limit where the agent population is composed only by fundamentalists or only by chartists. For these limits we can derive some analitycal results which are also useful to clarify various properties of the general case with both strategies. Morever this analysis will also suggest an approximated solution of the general case which will be discussed in section 5.\\
Our discussion starts from the equation for price evolution (paper I \cite{paperoI} Eqs. (8) and (9)) that can be written as
\begin{equation}\label{price2}
p_{t+1}-p_{t} = x\, \frac{b}{M-1}(p_{t}-p_M)+ (1-x)\gamma(p_f-p_t) + \sigma \xi_t.
\end{equation}
For the sake of simplicity we have dropped the explicit temporal depedence of the moving average $p_M$ and of $x$ in Eq. (\ref{price2}).\\
In Eq. (\ref{price2}) the temporal scale is the elementary one which corresponds to a time increment $\Delta t=1$. Our analitycal results will refer to this unitary time scale. On the other hand, computational results will be estimated from returns calculated each $100$ time steps for convenience with respect to real data. Unless specifically stated the time scale of our discussion will the elementary one. The analitical formulas will be given in term of this elementary scale and in order to compare them with results on a scale which is $k$ times the elementary increments, it will be sufficient to make the following substitution
\begin{equation}
t\,\,\rightarrow\,\,t'=kt.
\end{equation}
In this way the new function of the new variable $t'$ will be expressed in time units of length $k\Delta t$.
\subsection{The case $x = 0$}
\begin{figure}[!t]
\begin{center}
\includegraphics[scale=0.40]{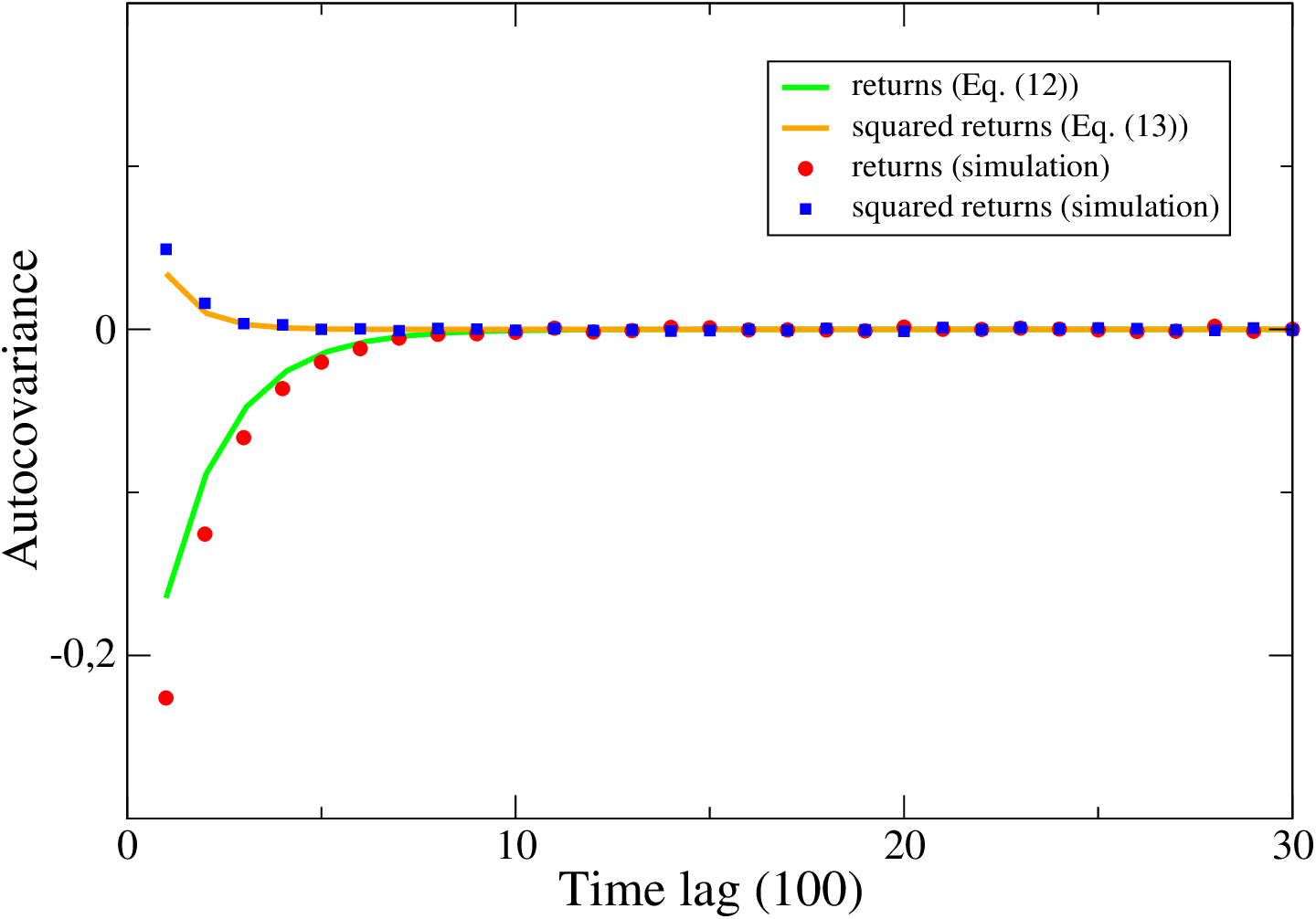}
\caption{The plot shows the comparison of Eq. \ref{efficiencyf} and Eq. \ref{volcluf} (solid lines) with the results of a simulation ($\circ$) and ($\Box$). We see that the autocovarinace function of returns,  which is a proxy to test the market efficiency, is negative when $ x =0$, conversely the volatility has a positive correlation but it is negligeable. In fact both of autocovariance functions have a decay time which is $\gamma^{-1}$ or fraction of $\gamma^{-1}$: in terms of efficiency this case is locally inefficient, instead in term of volatilty clustering this case substantially does not exhibit this phenomenon.}
\label{arbalpha0}
\end{center}
\end{figure}

In this case Eq. (\ref{price2}) reduces to:
\begin{equation}\label{pricefond}
p_{t+1}-p_{t} =\gamma(p_f-p_t) + \sigma\xi_t
\end{equation}
and, in order to avoid a trivial divergence of $p_t$, we consider the interval $0<\gamma<1$.\\ 
Equation (\ref{price2}) describes the path of a random walker subject to a force which tends to revert her to $p_f$ and $\gamma$ is the parameter of the strength of this force. In fact the first term of the right member of Eq. (\ref{price2}) can be written as the derivative (changed in sign) of the quadratic potential $(\gamma/2)\,(p_f-p_t)^2$ . \\
In the following calcutation we set $p_f=0$ wihout loss of generality since we are now considering the case of linear dynamics.
Given the initial value $p_0=0$ (without loss of generality), the explicit solution of Eq. (\ref{pricefond}) can be written at any time as
\begin{equation}\label{solutionf}
p_{t+1} = \sigma\sum_{j=0}^t\xi_j(1-\gamma)^{t-j}.
\end{equation}
We obviously have that\footnote{with $\textrm{E}[\cdot]$ we indicate the expectation value of the expression inside the squared brackets} $\textrm{E}[p_t] = 0$ and the second moment of the random variable $p_t$ takes the form:
\begin{equation}\label{second}
\textrm{E}(p_{t+1}^2)=\sigma^2 \bigg[\frac{(1-\gamma)^{-2}-(1-\gamma)^{2t}}{(1-\gamma)^{-2} -1}\bigg]
\end{equation}
It is sufficent to recall that
\begin{equation}\label{corrnulla}
E[\xi_i\,\xi_j] = \left\{
\begin{array}{rl}
0 & \text{if } i\neq j\\
1 & \text{if } i = j
\end{array} \right.
\end{equation}
in order to obtain Eq. (\ref{second}).\\ Equation (\ref{second}) shows that the second moment of $p_t$ is independent on time when $t\rightarrow\infty$. In fact it can be shown that the auto-regressive process (AR) that defines the case $ x =0$ is stationary. Hence in the following results of this section we assume that the limit  to the stationary regime has already been performed so that the temporal dependence can be omitted. Furthermore it is worth noticing that time-dependent terms exponentially go to zero with speed fixed by $\gamma$ which is typically $0.006$ in our model\footnote{thus the process is already nearly stationary for $t>1000$}.
\\We now turn our attention to the properties of the returns $r_t$ that we define as $r_{t+\Delta}= p_{t+\Delta}-p_t$ where $\Delta$ is the time lag at which we are studying the increments of the process (see appendix A for a discussion on the definition of the returns).
From Eq. (\ref{solutionf}) we find:
\begin{equation}\label{retexpli}
r_{t+\Delta} =\sigma(1-\gamma)^{t-1}\bigg[\sum_{j=0}^{t+\Delta-1}\xi_j(1-\gamma)^{\Delta-j}-\sum_{j=0}^{t-1}\xi_j(1-\gamma)^{-j}\bigg]
\end{equation}
and we can also compute the variance of returns (taking into account stationarity wa can write $r_{t+\Delta}\equiv r_\Delta$):
\begin{equation}\label{sigma_c}
E[r_\Delta^2] = \sigma^2 \frac{2(1-\gamma)^{-2}}{(1-\gamma)^{-2}-1}(1- (1-\gamma)^\Delta)
\end{equation}
that correctly reduces to the well known result for the RW $\sigma^2\Delta$ when $\gamma$ goes to zero. Equation (\ref{sigma_c}) can be calculated in a similar way of Eq. (\ref{second}) by Eq. (\ref{corrnulla}). The analysis of Eq. (\ref{sigma_c}) with respect to $\Delta$ leads to
\begin{equation}
E[r_\Delta^2] = \left\{
\begin{array}{rl}
\sigma^2 \frac{2(1-\gamma)^{-2}}{(1-\gamma)^{-2}-1}\gamma\Delta& \text{for small } \Delta\\
\sigma^2 \frac{2(1-\gamma)^{-2}}{(1-\gamma)^{-2}-1}\ & \text{for large } \Delta 
\end{array} \right.
\end{equation}
We see that for small values of $\Delta$ the behavior of Eq. (\ref{sigma_c}) is similar to one of the RW except for the size of fluctuations which are reduced by the attractive term due to the $p_f$. In the limit of large $\Delta$, instead, the fluctuations tend to a constant differently from the RW. \\ 
We now investigate the normalized autocovariance function of returns $\rho_{r_{\Delta}}(\tau)$ and of squared returns $\rho_{r_{\Delta}^2}(\tau)$ as a measure of market efficiency and of volatility clustering respectively (they depend only on the time difference $\tau$ because of the stationarity). We recall that $\rho_{r_{\Delta}}(\tau)$ and $\rho_{r_{\Delta}^2}(\tau)$ are defined as follows:
\begin{align}
\rho_{r_{\Delta}^2}(\tau)&=\frac{E[r_{\Delta,\tau}\,r_{\Delta,0}]-\mu_r^2}{\sigma_r^2}\label{correfffonf1}\\
\rho_{r_{\Delta}}(\tau)&=\frac{E[r_{\Delta,\tau}^2\,r_{\Delta,0}^2]-\sigma_r^4}{E[r_\Delta^4]}\label{corrvol2}
\end{align}
where $\sigma_r^2=E[r_\Delta^2]$ and $\mu_r=E[r_\Delta]$.
The former one can be computed in a similar way of previous fomulas because all the terms arising from Eq. (\ref{correfffonf1}) can be reduced to the evaluation of expressions like one in Eq. (\ref{corrnulla}). We find (it is a well-known results that the correlation function of an autoregressive model is characterized by an exponential decay):
\begin{equation}\label{efficiencyf}
\rho_{r_{\Delta}}(\tau) \simeq -\frac{\gamma\Delta}{2}e^{-\tau|\ln(1-\gamma)|} \stackrel{\gamma \ll 1}{\simeq}  -\frac{\gamma\Delta}{2}e^{-\tau\gamma}.
\end{equation}
Equation (\ref{efficiencyf}) shows the case $ x =0$ exhibits a negative correlation of price increments and this correlation has a slow decay rate since the characteristic time $\tau_0$ is about $\gamma^{-1}\approx 167$. This fact is not surprising because we have already noted that Eq. (\ref{pricefond}) describes a walker which is attracted by $p_f$. To put it simply, if we observe an increment of a certain sign at time $t$, increments of the opposite sign are more likely observed in the following time steps beacuse of the constraint of $p_f$.\\
In a similar way we can find from Eq. (\ref{corrvol2}):
\begin{equation}\label{volcluf}
\rho_{r_{\Delta}^2}(\tau)=f(\Delta,1-\gamma)e^{-\tau|\ln(1-\gamma)|} \stackrel{\gamma \ll 1}{\simeq} f(\Delta,1-\gamma)e^{-2\gamma\tau}
\end{equation}
where $f(\Delta,1-\gamma)$ is a function of the parameters $\Delta$ and $1-\gamma$ whose explicit form is rather complex but not very relevant since it refers to the values of the prefactor of the correlation function. As expected the decay characteristic time of the autocovariance function of squared returns is half the value found in Eq. (\ref{efficiencyf}) because the path of the price is a fractal with dimension $1/2$. But, while a decay time of about $\gamma^{-1}$ is rather long if we are studying the deviation from efficiency, a decay of $\gamma^{-1}/2$ is very short with respect to the real volatility clustering.\\
From a financial point of view these two results mean that there is a predictability of price increments for about 200 price increments (at the elementary scale) corrisponding to a rather long inefficiency while the volatility clustering decays quicker that ineffiency, so that the case $ x =0$ shows a behavior which is the opposite of reality.
In Fig. \ref{arbalpha0} we compare our analytical results with computer simulations of the case $ x =0$.
Before proceeding to the next section it is interesting to make some comments on the continous limit of the case $ x =0$. In this limit the price equation becomes ($p_f=0$):
\begin{equation}\label{OU}
dp(t)=-\gamma\, p(t)dt +\sigma dw_t
\end{equation}
where $dw_t$ is the differential of the Wiener process. The stochastic differential Eq. (\ref{OU}) is the well known Ornstein-Uhlenbeck process (for further details see \cite{Grimmett:2004lq}). The continous limit of Eq. (\ref{solutionf}) statisfying the same intial condition is:
\begin{equation}\label{OU2}
p(t)=\sigma e^{-\gamma t}\,\int_0^t\,e^{\gamma s}\,dw_s
\end{equation}
where the integral of the right side of Eq. (\ref{OU2}) is a It\=o stochastic integral. In order to test the results of the discrete time case we can compute the expressions corresponding to Eqs. (\ref{sigma_c}) and (\ref{efficiencyf}) in the continuous limit. We have verified that the formulas of the discrete case correctly reduce to the continuous one in the appropiate limit of the parameters\\

\subsection{The case $x = 1$}
\begin{figure}[htbp]
\begin{center}
\includegraphics[scale=0.4]{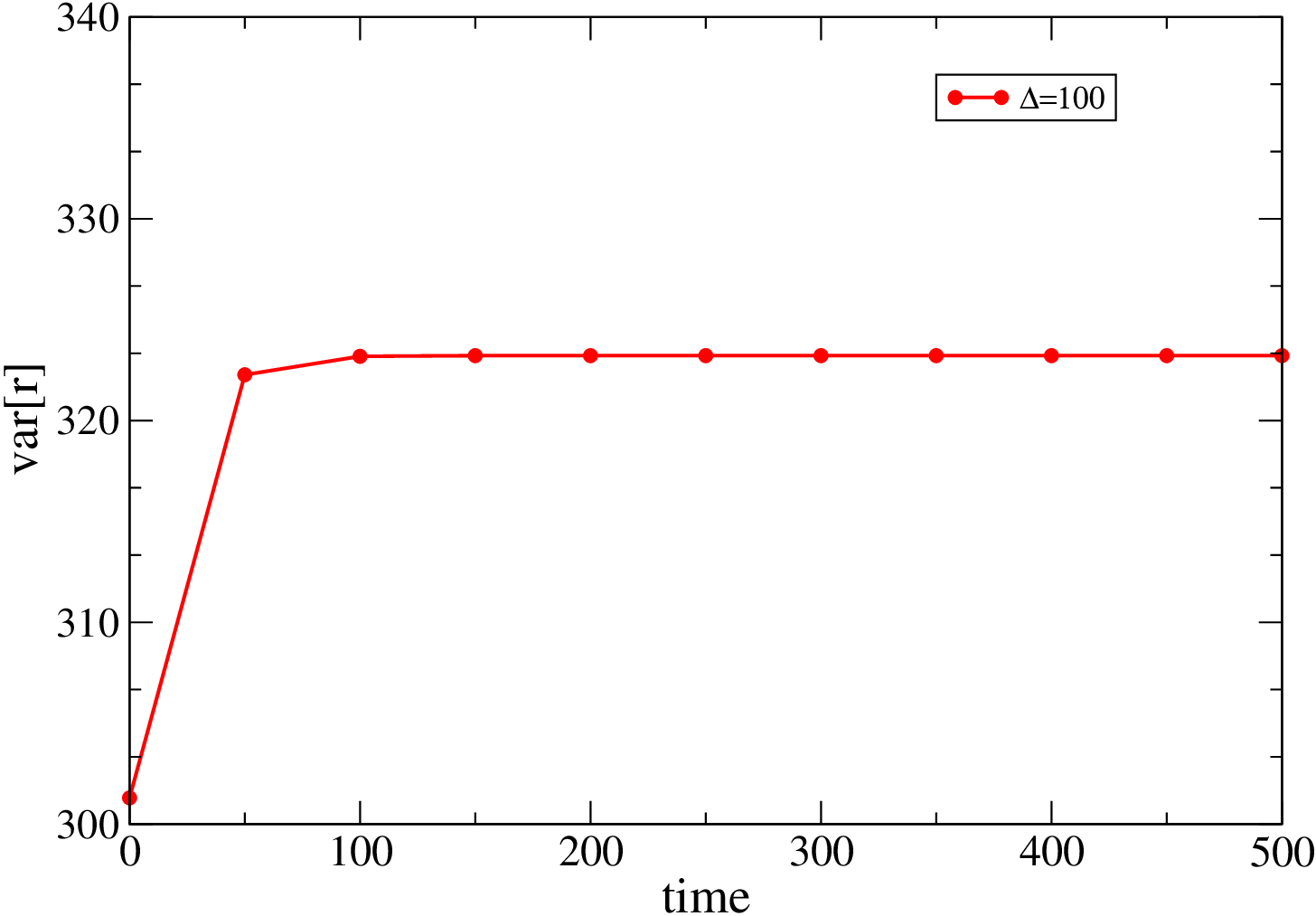}
\caption{In this picture we show the quick convergence to the stationary value of the variance of returns for time lags equal to $100$. The existence of the stationary limit of Eq. (\ref{retchartist}) is also supported by the fact that all eigenvalues of the matrix $\mathbf{T}$ are less than $1$.}
\label{convervar}
\end{center}
\end{figure}

\begin{figure}[htbp]
\begin{center}
\includegraphics[scale=0.40]{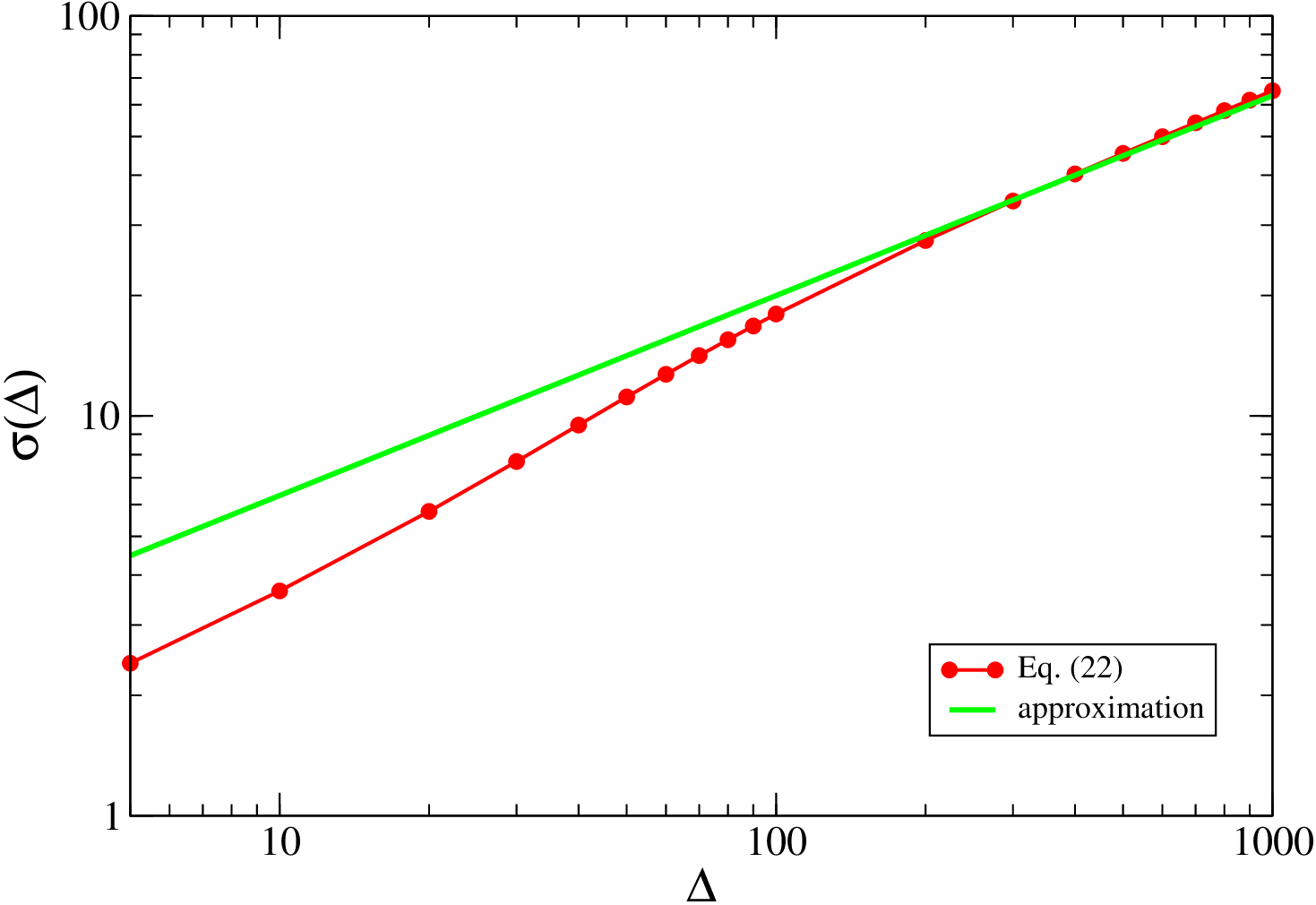}
\caption{Eq. (\ref{retchartist}), which is the exact expression of the variance or returns in the case $x=1$, predicts the superdiffusion of the process in the case $x=1$ for small value of time lag $\Delta$, in fact the slope of the red line is larger that the slope of the green solid line which behaves like a random walk.  As observed in \cite{taka1,taka2} the approximation proposed by these authors works very well when $\Delta\gg100$ where the process is close to a random walk except for a magnification of the effective $\sigma(0)$. }
\label{taka}
\end{center}
\end{figure}

\begin{figure}[htbp]
\begin{center}
\includegraphics[scale=0.40]{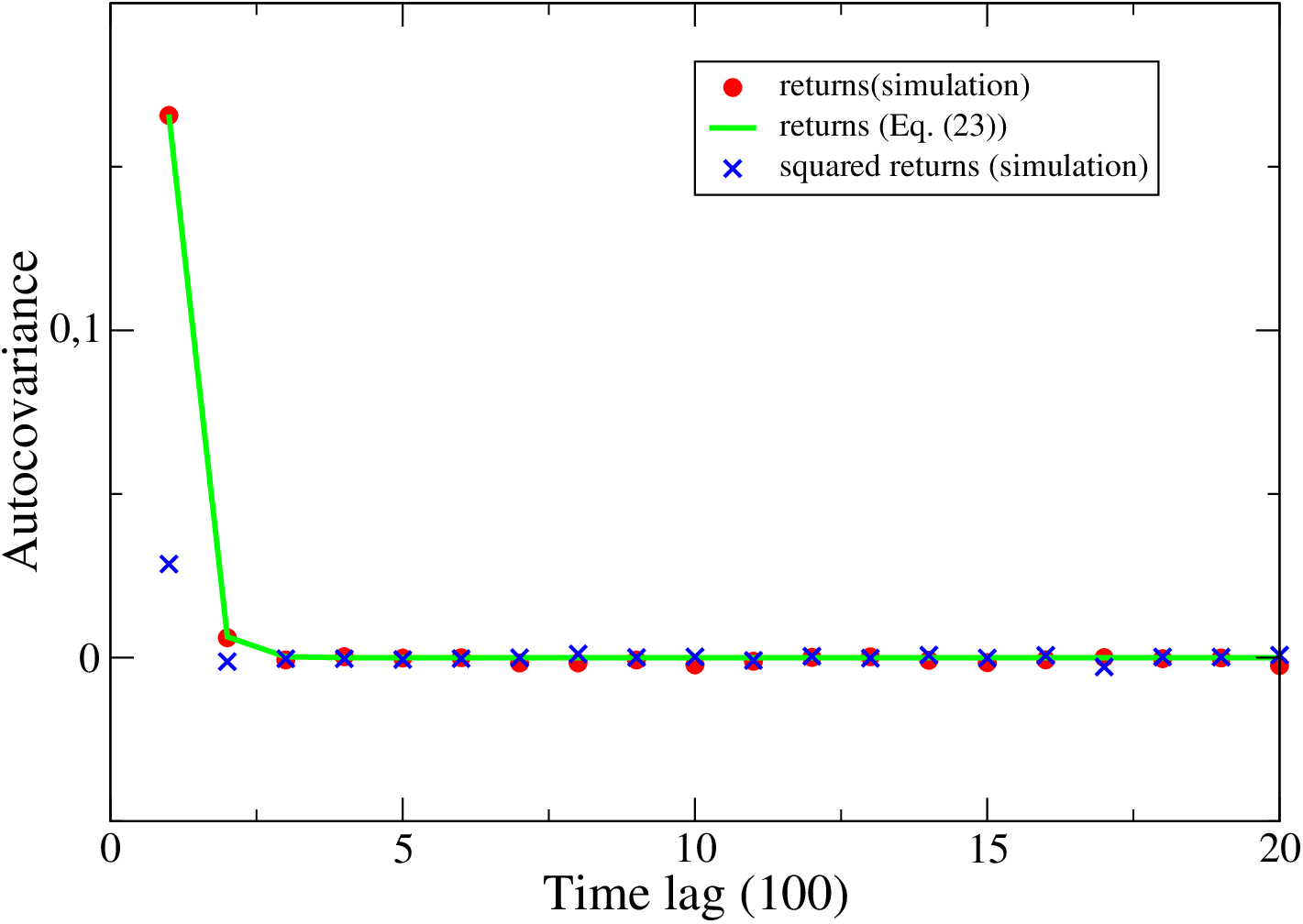}
\caption{As in the case $x=0$ we show how and if  the case $x=1$ reproduces the Stylised Facts, this case is characherized by a positive inefficiency whose temporal length is similar to the case of $x=0$, instead once again we see that neither the moving average is able to reproduce a significant and persistent volatility clustering}
\label{chartists}
\end{center}
\end{figure}

The development of a bubble or a crash corresponds to a strong increase of the fraction of the chartists, here we are going to explore the limiting case in which all agents are chartists. In such a case $x=1$ Eq. (\ref{price2}) becomes:
\begin{equation}\label{pricec}
p_{t+1}-p_{t} = \frac{b}{M-1}(p_{t}-p_M)+  \sigma\xi_t
\end{equation}
This equation describes a random walk in a quadratic field potential as the previous case but this time the force is repulsive and the istantaneous (unstable) equilibrium position is the value of the moving average $p_M$.\\
This approach to model the stochastic process of the price was first introduced by \cite{taka1,taka2} and following these papers we rewrite Eq. (\ref{pricec}) as a telescoping sum:
\begin{equation}\label{telescoping}
p_{t+1}-p_{t} =  a \sum_{i=1}^{M}(M+1-i)(p_{t-i+1}-p_{t-i})+  \sigma\xi_t
\end{equation}
where $a=b/(M(M-1))$. 
In this way we can find a simple recursive formula for a M-dimensional system whose first component is equivalent to Eq. (\ref{telescoping}):
\begin{equation}\label{vect}
\vec{r}_{t+1}=\mathbf{T}\,\vec{r}_t+\vec{\eta}_t
\end{equation}
where $\mathbf{T}$ is a $M\times M$ matrix defined as:
\begin{equation}
\mathbf{T}=\left(
\begin{array}{ccccc}
  a_1&  a_2 &\ldots& \ldots & a_M   \\
  1&0   &\ldots& \ldots & 0  \\
 0 & 1  & \ldots & \ldots &0\\
 \vdots&\vdots&\ddots & \ddots&\vdots\\
   0 & 0  & \ldots&1&0
\end{array}
\right)\qquad a_k=a(M+1-k)
\end{equation}
and $\vec{r}_t$ and $\vec{\eta}_t$ are vectors M-dimensional so defined:
\begin{equation}
\vec{r}_t=
\left(
\begin{array}{c}
p_t -p_{t-1}  \\
p_{t-1} -p_{t-2}    \\
 \vdots\\
p_{t-M+1}-p_{t-M}    
\end{array}
\right)
\qquad\qquad
\vec{\eta}_t=
\left(
\begin{array}{c}
\sigma\xi_t \\
0   \\
 \vdots\\
0  
\end{array}
\right)
\end{equation}
The authors of \cite{taka1,taka2} propose a simple approximation to calculate the variance of returns and find that the variance of the RW is magnified by a b-dependent factor larger than 1 so that the repulsive potential magnifies fluctuations with respects to a RW. We note that it is possible to solve the vectorial Eq. (\ref{vect}) recoursively:
\begin{equation}\label{solchar}
\vec{r}_{t+\Delta}=(\sum_{k=0}^{\Delta}\mathbf{T}^{t+k})\vec{y}_0+\sum_{k=0}^{t}(\sum_{h=0}^{\Delta-1}\mathbf{T}^{t+h-k})\vec{\eta}_k + \sum_{k=1}^{\Delta-1}(\sum_{h=1}^{\Delta-k}\mathbf{T}^{h-1})\vec{\eta}_{t+k}
\end{equation}
where $\vec{y}_0$ is the initial condition, for the sake of simplicity we assume that $y_{0}^i = 0$ for $i=1,\ldots,M$.
The solution of Eq. (\ref{pricec}) can easily be recovered by considering the first component $r_t^1$ of the vector $\vec{r}_t$. We find that $E[r_t^1]=0$ and the variance is:
\begin{equation}\label{retchartist}
E[(r_{t+\Delta}^1)^2] = \sigma^2\bigg[\sum_{k=0}^t( x _{k,t})^2 + \sum_{k=1}^{\Delta-1}(\beta_{k})^2 \bigg]
\end{equation}
where $ x _{k,t}=(\sum_{h=0}^{\Delta-1}\mathbf{T}^{t+h-k})_{11}$ and $\beta_{k}=(\sum_{h=1}^{\Delta-k}\mathbf{T}^{h-1})_{11}$. The structure of Eq. (\ref{retchartist}) is rather interesting, in fact it is composed of two sums: the first one depends explicitely on time $t$, while the second depends only on $\Delta$. This is because it originates from the last term of the right member of Eq. (\ref{solchar}) that depends on $t$ only through the index of the gaussian random variable $\xi_t$. Therefore this dependence will eventually introduces only a Kronecker's Delta function if some expectation values are calculated. In order to have a stationary state the Eq. (\ref{retchartist}) can no more depend on $t$ and indeed in Fig. \ref{convervar} we have verified that the process appears to converge to a stationary value for the variance when $t$ grows.\\
As counter check (and as a more formal test to verify the existence of the stationary state) we can evaluate numerically the spectrum of the matrix $\mathbf{T}$. Naming $\lambda_i$ with $i=1,\ldots,M$ the eigenvalues, we find that $\mathbb{R}(\lambda_i)<1$ $\forall i$ and thus the time dependent term of Eq. (\ref{solchar}) will behave like a geometric progression with ratio less than $1$, that is the sums are convergent. So, as in case $ x =0$, the process admits an asymptotic stationary state and in practice this means that the process is already stationary for $t>500$.\\
In Fig. \ref{taka} instead we compare our exact result with the approximation of the return variance given in \cite{taka1,taka2}, the approximated formula gives a very good agreement in the asymptotic limit where the process is substantialy a RW. On the other hand the approximation does not reproduce the super diffusion for $\Delta <1000$. In fact at this time scale the repulsive role played by the moving average is crucial.\\
In a similar way we can calculate the normalized autocovariance function of returns $\rho_{r_{\Delta}}(\tau)$:
\begin{equation}\label{autoretchar}
\rho_{r_{\Delta}}(\tau)=\lim_{t\rightarrow\infty} \frac{\sum_{k=0}^{t} x _{k,t} x _{k,t+\Delta}+\sum_{k=1}^{\Delta-1}\beta_{k} x _{k+k,t+\Delta}}{\sum_{k=0}^t( x _{k,t})^2 + \sum_{k=1}^{\Delta-1}(\beta_{k})^2}
\end{equation}
The numerator of Eq. (\ref{autoretchar}) is positive-definite (and then $\rho_{r_{\Delta}}(\tau)$ too) because it is a linear combination of products of positive terms. In Fig. \ref{chartists} we report the estimation of $\rho_{r_{\Delta}}(\tau)$ from a numerical simulation of the process in the case $x=1$ compared to the analitical result of Eq. (\ref{autoretchar}).\\
The autocovariance of squared returns can be also computed in principle but its structure is very complex, in Fig. {chartists} the asterics corresponds to a simulation which shows its positive sign and the fast convergence to zero. Once again we have a process which is characterized by a quite long inefficiency (positive this time), and on the other hand by the absence of persistency of volatility clustering.
To summarize the studies of this section we see that the two extreme cases in which a type of agent dominates on the other one correponds to two different regimes. One of these are characterized either by gaussian fluctuations smaller than those observed in a pure RW and by negative arbitrage ($x=0$). The other one instead has gaussian fluctuations larger than those observed in a pure RW and positive arbitrage ($x=1$). Neither of these two regimes show volatility clustering, therefore the next question is how these ingredients are mixed when we take into account the dynamics of $x$ so that it can assume values between 0 and 1. Next sections will answer this interesting question.
\clearpage
\section{Diffusional properties}
Diffusion processes are time-continuous Markov processes that have to satisfy three conditions (see \cite{stocastoca} for  a mathematical formulation of this characterization). These conditions, roughly speaking, state that diffusion processes are fully characterized by their drift (i.e. their mean) and by their diffusion coefficient (i.e. their variance). Diffusion processes are essentially gaussian processes which are fully specified by the knowledge of the mean and the variance\footnote{however a gaussian process can be non-Markov differently from a diffusion process which is necessary a Markov process}. \\
One should notice that these diffusional properties (limited to the squared returns) do not provide a complete information on the system in the presence of anomalous tails, however we are going to see that the diffusion analysis leads in any case to some interesting information. 
Before moving to the statistical analysis we would like to point out that the process in the case $x=0$ is gaussian and Markovian. Also the case $x=1$ is gaussian, being linear in the price and in the white noise, but no more Markovian due to  the moving average which introduces a term that explicitly depends on the price path. The full process (Eq. (\ref{price2})) would be gaussian if $x$ was fixed because of the linearity of Eq. (\ref{price2}) (but non Markovian). Nevertheless the random process which drives the switch of populations destroy the gaussianity and the return pdf actually has non gaussian tails. We will show this effect in section 5 where we propose an approximated mechanism to explain the non gaussian tails.\\

\begin{figure}[htbp]
\begin{center}
\includegraphics[scale=0.40]{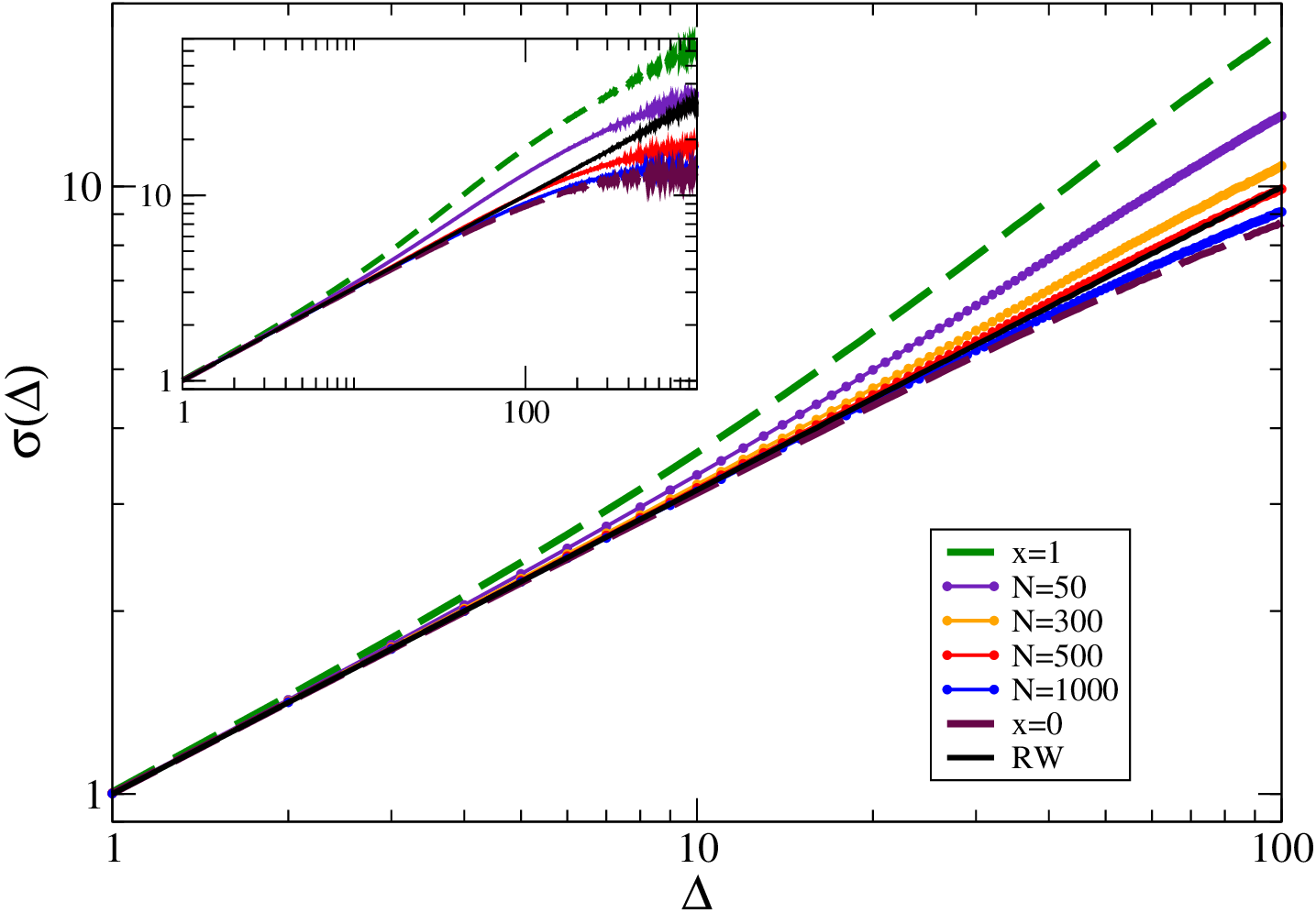}
\caption{Investigation of the scaling behavior of the second moment of the returns, that is the diffusional properties, for $x=0;1$ and for different values of $N$ and in the range of small values of $\Delta$. In the main plot we see that the short scale behavior is approximately of the type $\Delta^{\mu}$ and the exponent $\mu$ varies from the maximum $0.7$ when $x=1$ (superdiffusion) to the minimum $0.44$ when $x=0$ (subdiffusion). When $N\approx 500$ the behavior converges to one of a random walk. The small insert shows instead the begining of the regime dominated by fundamentalists which develops in the limit of large $\Delta$.}
\label{diffushort}
\end{center}
\end{figure}

\begin{figure}[htbp]
\begin{center}
\includegraphics[scale=0.40]{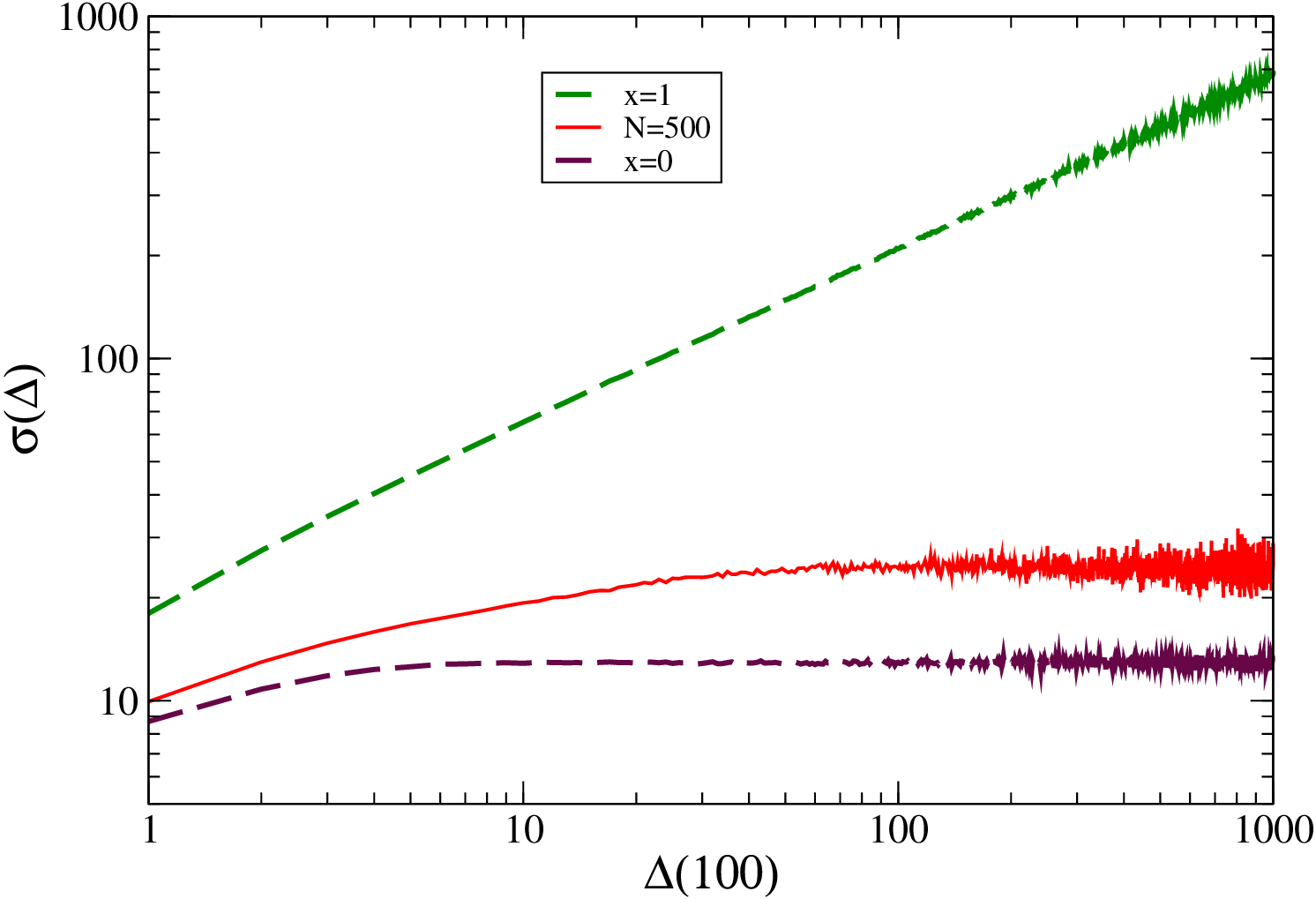}
\caption{Scaling behavior of the second moment of returns. Diffusional properties for $x=0;1$ and for the full mixed process with $N=500$ in the limit $\Delta\rightarrow\infty$. While at the short time scale the subdiffusion of the case $x =0$ and the superdiffusion of the case $x=1$ someway balance out, when $\Delta$ grows the variance tends to a constant as in the case $x=0$.}
\label{diffulong}
\end{center}
\end{figure}

It is well-known that the variance of the increments of a RW is proportional to the time scale $\Delta$ that is $\sigma(\Delta)=\sigma_0\Delta^{\mu_{rw}}$ with $\mu_{rw}=0.5$. We now investigate the scaling properties of the variance of increments of the process of Eq. (\ref{price2}) as a function of the time lag $\Delta$ at which returns are calculated for different values of $N$ and we sometimes find deviations from $\mu_{rw}$. We also investigate the variance of the increments on two different temporal windows, the former could be defined a short-scale analysis, the latter instead an analysis on long temporal horizons.\\
\paragraph{Short time scale}
We report the plot of the variance pattern in Fig. \ref{diffushort} while in table \ref{tabmu} we show the corresponding results of a fit with a power law of the curves of Fig. \ref{diffushort} in the short scale case \footnote{the evaluation of the error bars when we deal with dependent variables is longly debated so we prefer dropping them and consequently we propose the fitted parameters without indication of their statistical errors}.\\ 
Once again we stress the crucial role played by the number of agents $N$ with respect to the statistical properties of the model. In fact we clearly see in table \ref{tabmu} that the variance, for a fixed $\Delta$, is a decreasing function of $N$. This can be understood in term of the population dynamics  (see section 1 and \cite{paperoI,paperNP}: when the number of agents $N$ increases, the fraction $x$ of chartists tends to zero because of the exponential term of the asymptopic stable distribution of Eq. (21) of paper I \cite{paperoI} and chartists have previously been recognized as fluctuation magnifiers, while fundamentalists as fluctuation reducers.\\
The most striking result is however the value of $\mu$ found for $N\approx 500$, when it is nearly equal to $\mu_{rw}$. The increments have the same scaling behavior of a RW but it would be misleading to conclude our process is a RW because we know that the process is non gaussian and non Markov since it exhibits non trivial correlations and non gaussian tails.This implies that a complex intermittent behavior may not be detected by a simple diffusion analysis. It is worth noting that the behavior for any $N$ is bounded by the limitating cases $x=0$ and $x=1$.\\ The apparent RW-like scaling behavior which for our parameters occur around $N=500$, is the results of the a balance of the opposite tendencies due to the fundamentalists and chartists. A similar balancing situation has been discussed in \cite{taka1,taka2} even though the framework is rather different than the present one.\clearpage
\begin{table}[!htbp]
\begin{center}
\begin{tabular}{|c|c||c|c|}
\hline
N &$\mu$ &N & $\mu$\\ \hline
 RW & 0.500 &300 & 0.523\\
$x=0$ & 0.439&400 & 0.509\\ 
$x=1$ & 0.701&500 & 0.496\\ 
10 & 0.609&600 & 0.482\\
50 & 0.596&750 & 0.466\\
100 & 0.581&1000 & 0.455\\\
200 & 0.555&5000 & 0.440\\
250 & 0.541& &\\ \hline
\end{tabular}
\end{center}
\caption{Results of fit of the exponent $\mu$ for $\Delta \in [1,100]$}
\label{tabmu}
\end{table}
\clearpage
\paragraph{Long time scale}
When we turn our attention to a longer temporal horizon ($\Delta \gg 100$), we can observe very different features (see Fig. \ref{diffulong}). According to Eq. (\ref{sigma_c}) the variance of the case $x=0$ goes to a constant when $\Delta \rightarrow\infty$ while from Eq. (\ref{taka}) the variance of $x=1$ grows as in a RW when $\Delta \rightarrow\infty$ ($\mu_{x=1}$ is larger than 0.5 at short times and asymptotically tends to 0.5 from above). \\
The balancing effect fails in this limit (in Fig. \ref{diffulong} we show only the case $N=500$ because for all $N$ the variance tends to a constant but this constant value depends on $N$). In order to explain what happens when $\Delta$ grows we observe that the characteristic time scale of the moving average is M (or few times M). Therefore, on time-scale larger than M, a coarse-grained version of the process relate to $p_M$ would be nearly undistinguishable from a RW. Instead the case $x=0$ does not exhibit diffusional properties since
\begin{equation}
c_\infty = \lim_{\Delta\rightarrow\infty}\frac{\sigma^2(\Delta)}{\Delta}=0
\end{equation}
where $c_\infty$ is the asymptotic diffusion coefficient.
In this way if $\Delta$ becomes very large (i.e. the coarse-grained limit) the term relative to the moving average behaves like an additional noise. In fact, as we have seen in Fig. \ref{diffulong}, we just observe a greater effective variance than the one predicted by Eq. (\ref{sigma_c}).\clearpage
\section{Transition to gaussianity?}
\begin{figure}[htbp]
\begin{center}
\includegraphics[scale=0.37]{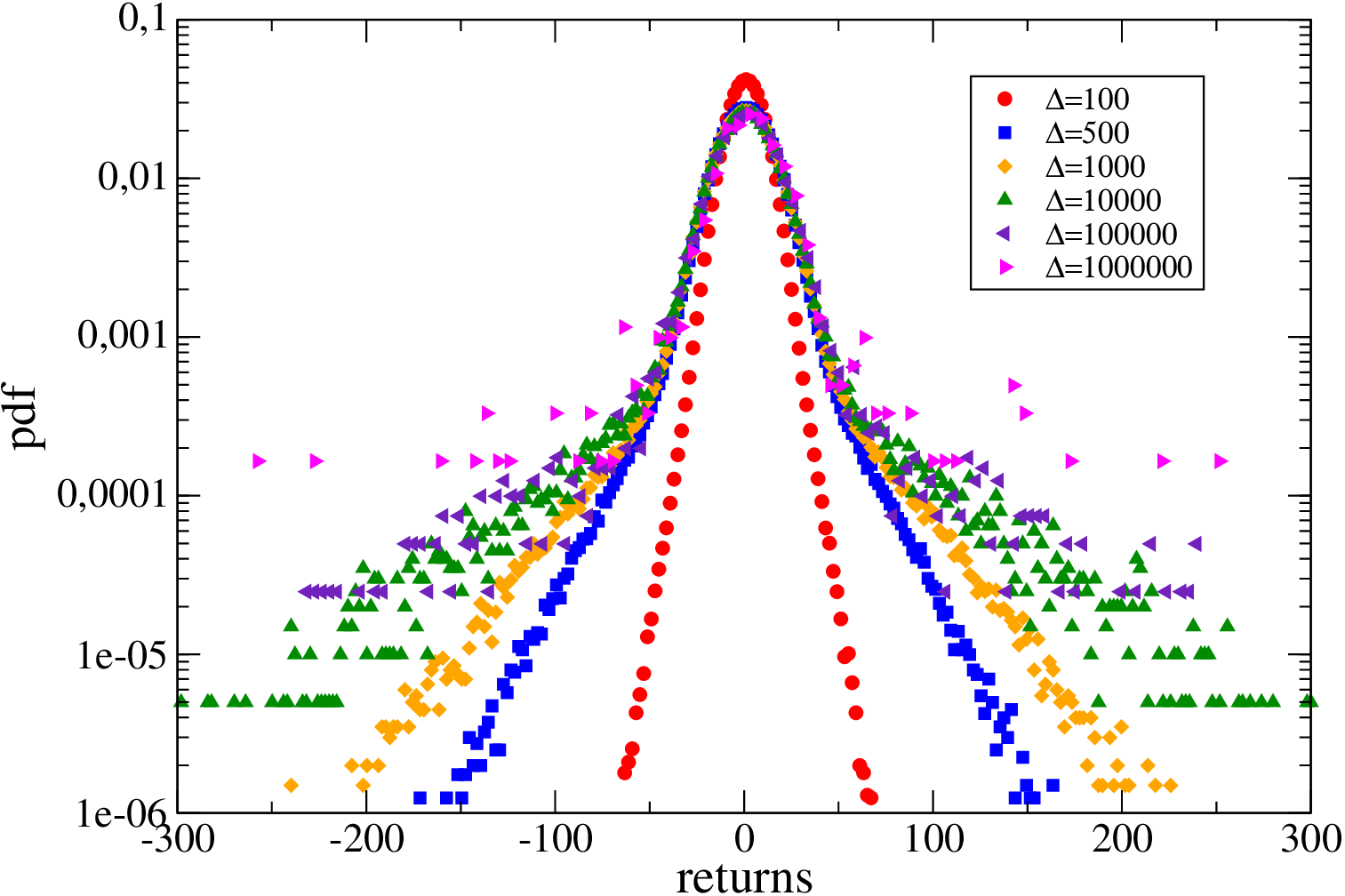}
\caption{Probability density function of returns for different values of time lag $\Delta$ at which returns are performed. Our minimal model does not seem to show a transition to a gaussian regime since the non gaussian tails persists even for very large values of $\Delta$.}
\label{allpdf1}
\end{center}
\end{figure}

\begin{figure}[htbp]
\begin{center}
\includegraphics[scale=0.37]{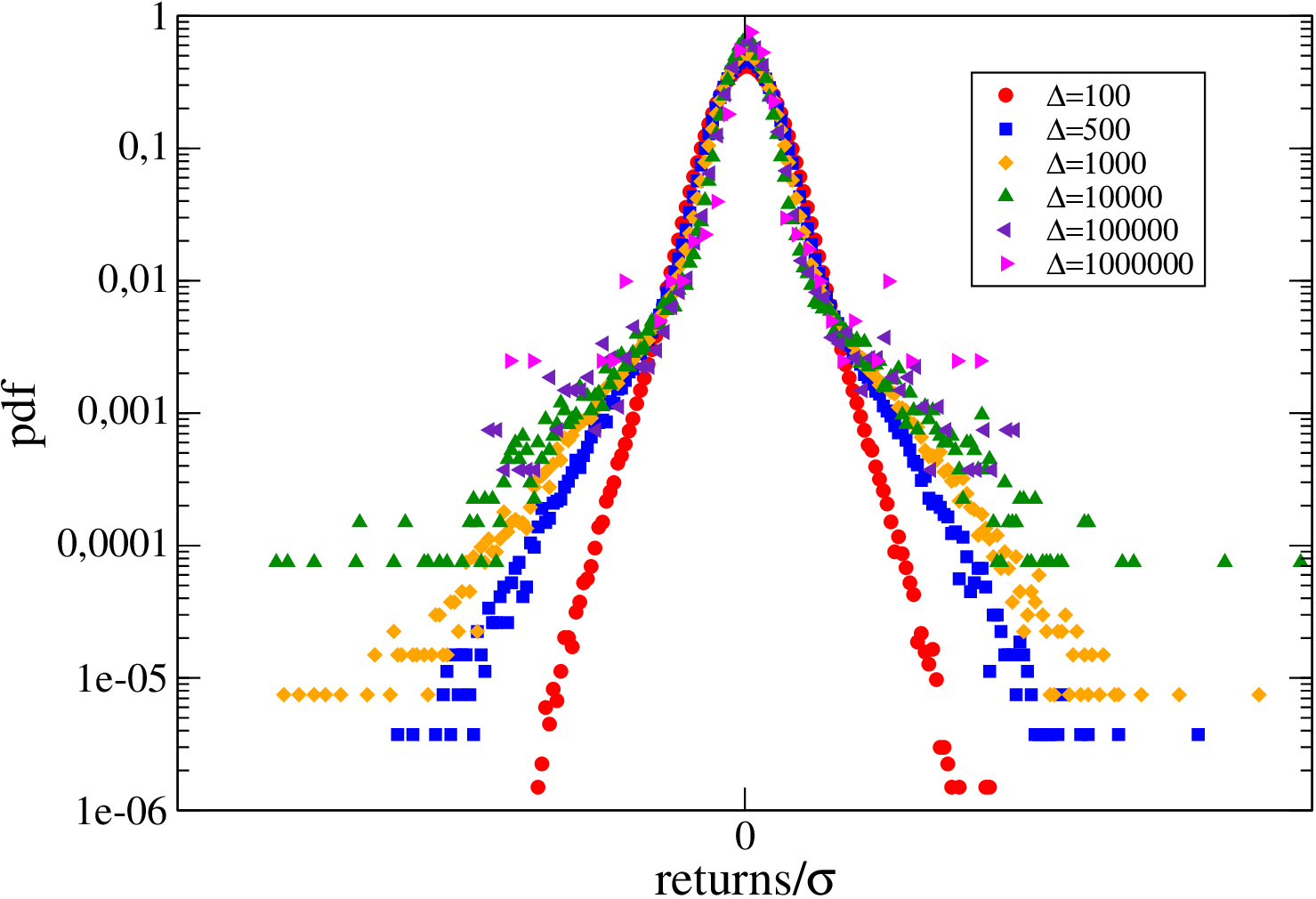}
\caption{Probability density function of returns normalized by their variance. The non collapsing tails suggest that the scaling behavior of the variance captures only partially the process. We will try a different normalization of returns in Appendix C but we will not obtain collapsing tails for any $\Delta$. These difficulties arise from the fact that tails draw origin from the non linearity introduced by the population dynamics.}
\label{allpdf2}
\end{center}
\end{figure}

\begin{figure}[htbp]
\begin{center}
\includegraphics[scale=0.60]{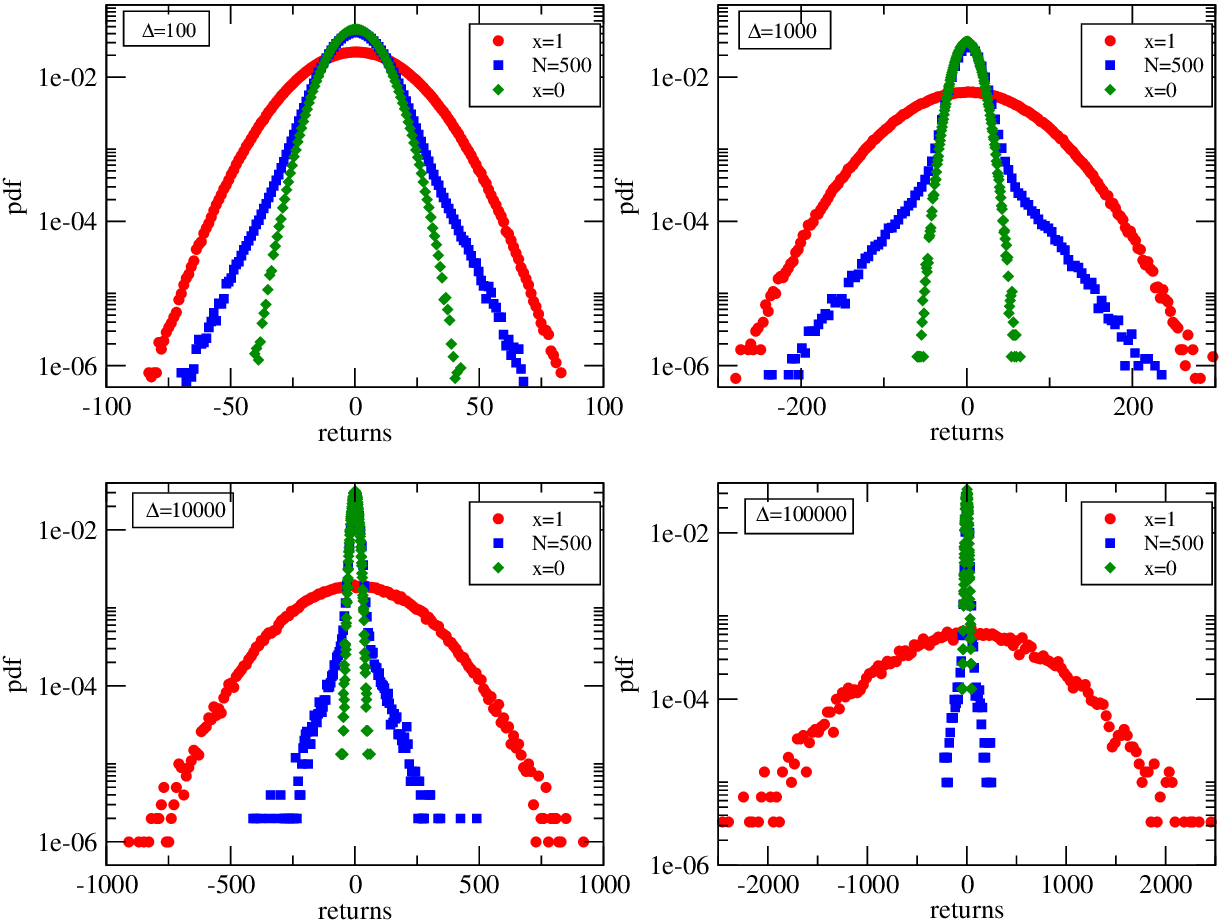}
\caption{The four panels show the comparison of the return probability density function for the extreme cases $x =0;1$ (which are gaussians) and for the mixed dynamics for $N=500$ for different values of $\Delta$. The growing gaussian of the case $x =1$ seems to stretch the non gaussian tails which are produced by a non trivial interaction of these two gaussians through the population dynamics, that is through by the temporal dependence of $ x $.}
\label{pdfCFN}
\end{center}
\end{figure}

In the previous section we have studied the $\Delta-$dependence of the variance, but the parameter $\Delta$ may a priori play a role in the shape of the return pdf too. The motivation of such an interest can be traced in many analysis of market data (see \cite{Cont:2001, lebaronbreve}) where data appear to be well-fitted by a gaussian when $\Delta\rightarrow\infty$. This phenomenon is still unclear, some authors invoke the central limit theorem (CLT) so that the validity of CLT, that fails for small value of $\Delta$, seems to be recovered in this limit although it is still unclear if CLT or critical-phenomena-like arguments should be used or neither of them.\\
\\
We are going to discuss in detail the problem of the Fat Tails within our model. 
In Fig. \ref{allpdf1} we display the pdf of the returns \footnote{data are from a very long simulation $t=10^9$} while in Fig. \ref{allpdf2} we show the same results but we rescale the returns dividing them by their variance. From these figures it is evident that in our model Fat Tails do not dissapear for $\Delta \rightarrow\infty$, this may appear contradictory with the results of Fig. \ref{diffulong} where the variance is shown to converge to a finite value for $\Delta \rightarrow\infty$. We are going to see that the origin of this apparent discrepancy is due to the fact that the second moment is not appropriate to discuss the Fat Tails because it is dominated by the central part of the pdf.\\ In Fig. \ref{allpdf1} and {allpd2} we see that all the pdfs have the same behavior: in the central region they appear to follow the shape of a gaussian while, in the region of rare events, pdfs have tails decaying slower than gaussian. Therefore the interaction between chartists and fundamentalists must be the key element to increase the probability of observing large fluctuations. An analysis of the data shows that the central gaussian behavior is essential due to the action of fundamentalists ($x=0$) as discussed before.\\
The fact that the rescaled return pdfs do not collapse everywhere in the same curve confirms that the scaling property of the variance does not fully characterize the process and thus the rescaled returns exhibits only a partial collapse that indeed fails in the tail region (see Appendix C).\\
\\
In order to answer why in this mininal ABM deviations from gaussianity do not disappear when $\Delta\rightarrow\infty$ we have to consider the microscopic mechanism which leads to the tails. \\
On this account Fig. \ref{pdfCFN} is very instructive: we show the return pdf for $N=500$ and the corrisponding pdf of the cases $x=0;1$ for the same $\Delta$. We see that the central part of the curve for $N=500$ (blue) is very close to the gaussian of the case $x=0$, while the Fat Tails appear to be somehow caused by the increasing width of the gaussian of the case $x=1$. Since we have already seen that the model would be gaussian if $x$ was fixed, we can conclude the non-linearity deriving from the dynamics of $x$ must be the key to understand the emergence of non gaussian tails.\\
The possible relation of our results to real data should be considered with great care. For example our model is tuned to elucidate the origin of the SF and for simplicity we have set $p_f$ to be a fixed value independent on time. This is useful from a conceptual point of view but it is certainly non realistic assumption. As we are going to see in section 6 the introduction of a RW dynamics for $p_f(t)$ would change completely the asymptotic behavior of the variance and would lead to a real crossover to gaussian behavior at long time.
\section{Nature of Fat Tails}
\begin{figure}[htbp]
\begin{center}
\includegraphics[scale=0.40]{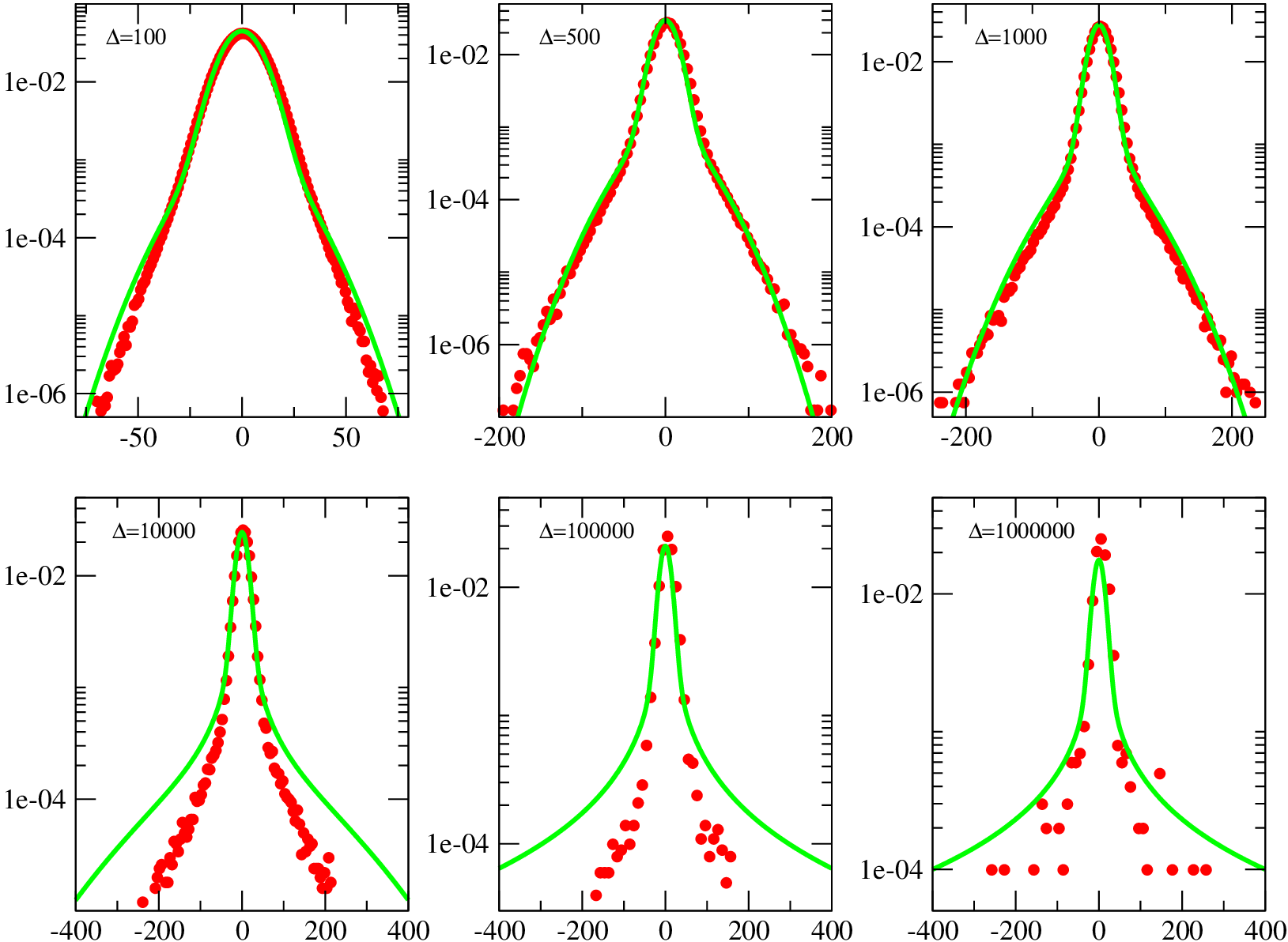}
\caption{Comparison of the approximation proposed in Eq. (\ref{mediaalpha}) with the probability density function of returns estimated from a simulation for different values of $\Delta$. The approximation gives a good agreement with simulation data until $\Delta<1000$. For larger values of time lag $\Delta$ the tails are overestimated by the approximation. This corresponds to an overestimation of the effective fraction of chartists when the average on $x$ is performed. This is in agreement with the behavior of the variance shown in Fig. \ref{diffulong}.}
\label{approxpdf}
\end{center}
\end{figure}

\begin{figure}[htbp]
\begin{center}
\includegraphics[scale=0.40]{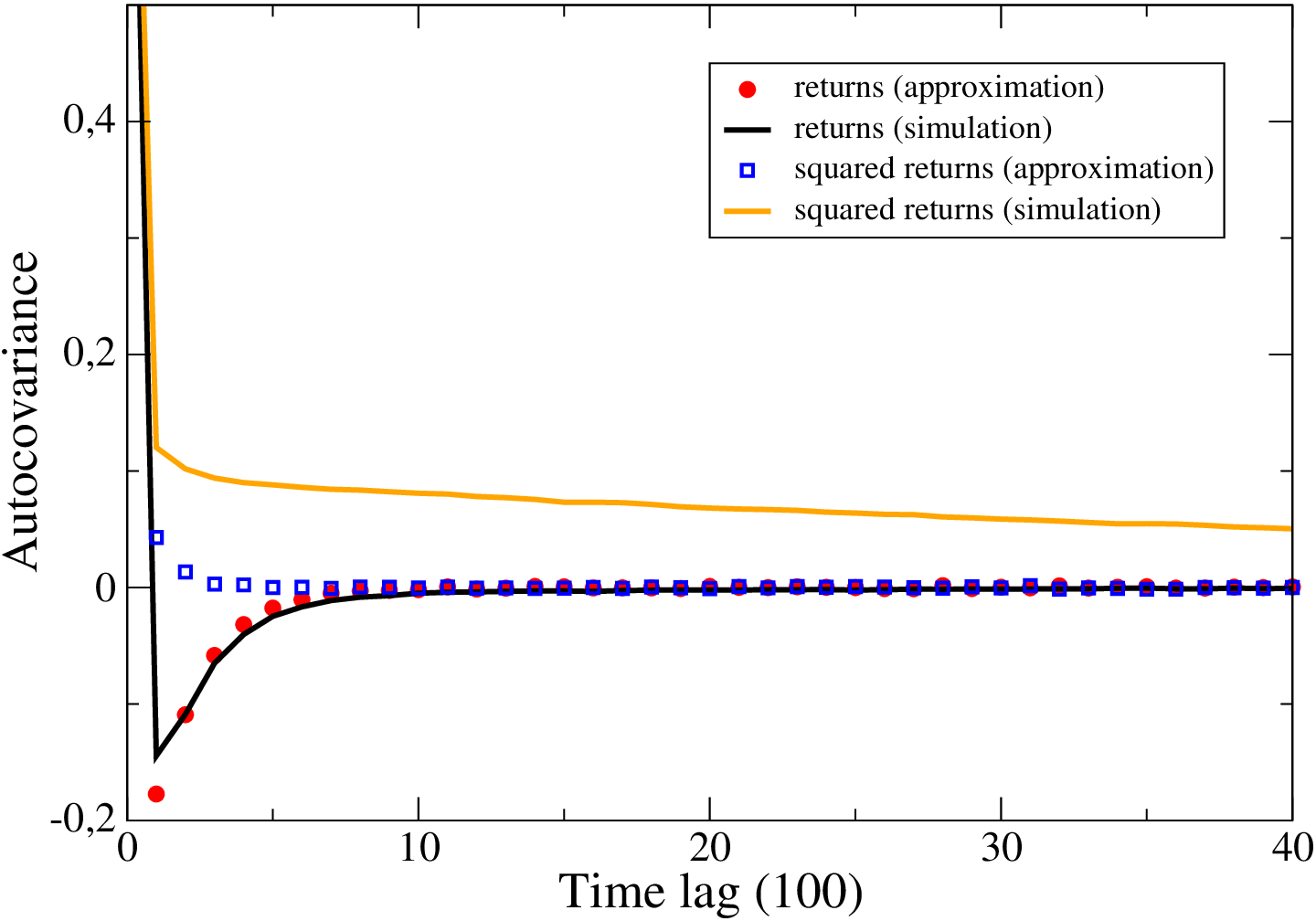}
\caption{The simple approximation proposed in this section also allows to extimate the autocovariance of returns and of squared returns. The global efficiency of the model seems to be well reproduced by the simple superposition of the solution of the case $x=0;1$. The approximation fails in reproducing volatility clustering (even for $t<1000$) because the individual cases $x=0;1$ do not reproduce this Stylised Fact. This comparison clarifies the key role of the population dynamics (i.e. of $x$) in order to generate bubbles which gives non-zero correlation for volatility.}
\label{approxarb}
\end{center}
\end{figure}

We now propose a simple approximation which clarifies the origin of the non gaussian tails.
In section 2 we have found that the return pdf of the case $x=0$ and $x=1$ are respectively 
\begin{align}
g_f(y):=g(y|x=0)=\frac{1}{\sqrt{2\pi\sigma_f^2}}e^{-\frac{y^2}{2\sigma_f^2}}\label{pdff}\\
g_c(y):=g(y|x=1)=\frac{1}{\sqrt{2\pi\sigma_c^2(\Delta)}}e^{-\frac{y^2}{2\sigma_c^2(\Delta)}}\label{pdfc}
\end{align}
with $\sigma_f^2\approx\, \textrm{cost}$ when $\Delta \gg 1$ and $\sigma_c^2(\Delta)\approx \sigma_{c,0}^2\Delta$.
In other words we know the solution of our model in this two limiting cases. It is instructive to make the simple assumption that the general solution $r_N$ for a certain N can be written as a superposition of the two cases
\begin{equation}\label{rapp}
r_{N}=x r_c+(1-x)r_f
\end{equation}
where $r_f$, $r_c$ are the returns of the case $x=0$ and $x=1$ distributed respectively as Eq. (\ref{pdff}) and Eq. (\ref{pdfc}) and we assume stationarity. Hence the pdf of $r_N$, for a fixed valaue of $x$, is given by the following convolution
\begin{equation}
p(r|x)=\iint dydz \,g_{c,f}(y,z)\,\Delta\big(r-x y-(1-x)z\big)
\end{equation}
where $g_{c,f}(y,z)$ is the joint pdf of chartists and fundamentalists.\\We neglect the interactions between chartists and fundamentalists and in this approximation we can factorize $g_{c,f}(y,z)=g_f(y)g_c(z)$ and in the end we find that
\begin{equation}\label{pdfretapp}
p(r|x) = \frac{1}{\sqrt{2\pi\hat{\sigma}^2}}e^{-\frac{r^2}{2\hat{\sigma}^2}}
\end{equation}
where $\hat{\sigma}^2=x^2 \sigma_c^2(\Delta)+(1-x)^2\sigma_f^2$. This result is trivial because the convolution of two gaussian is a gaussian.\\However, a much more interesting result can be obtained by considering the equilibrium distribution $f_{eq}(x)$ of $x$, as given by Eq. (21) of I \cite{paperoI} and integrated Eq. (\ref{pdfretapp}) over this distribution of values of $x$. This integral can be numerically evaluated
\begin{equation}\label{mediaalpha}
E_x[p(r|x)]=\int_0^1dx\,f_{eq}(x)\,p(r|x).
\end{equation}
In Fig. \ref{approxpdf} we compare the approximated curve corresponding to Eq. (\ref{mediaalpha}) with the simulated one and we can see that there is a good agreement especially for $\Delta<10^4$.\\
\\
From Fig. \ref{approxpdf} one can see that, at short time, the superposition ($\Delta<1000$) is a very good approximation, but for longer time this approximation leads to an over-estimation of the tails. This can be undestood rather easily by considering, for example, the behavior discussed in relation of Fig. \ref{diffulong}. For long times the variance of the chartists alone diverges as in a normal diffusion process, on the other hand the variance of the fondamentalists alone is bounded to a fixed value. Considering the attractive role which $p_f$ plays in dynamics of the complete model it is clear that the variance of the complete model will not diverge. In this respect the incoherent superposition of the two limiting cases leads to a excess of variance with respect to the real dynamics of the model.  \\\\
This simple superposition approximation can also be used to test which features of the model depend crucially on the chartist-fundamentalist interaction.
Assuming that the general solution can be put in the form of Eq. (\ref{rapp}) and assuming that $r_c$ and $r_f$ are indenpendent and stationary processes, we can compute the approximated autocovariance function of returns $\rho_{r_{\Delta}}(\tau)$ and of squared returns $\rho_{r_{\Delta}^2}(\tau)$. The former is
\begin{equation}\label{arbapp}
\rho_{r_{\Delta}}(\tau|x) = \bigg[1+\frac{(1-x^2)}{x^2}\frac{\sigma_f^2}{\sigma_c^2}\bigg]^{-1}\rho_{r_{\Delta}^c}(\tau) + \bigg[1+\frac{x^2}{(1-x)^2}\frac{\sigma_c^2}{\sigma_f^2}\bigg]^{-1}\rho_{r_{\Delta}^f}(\tau)
\end{equation}
where $\sigma_f^2$ and $\rho_{r_{\Delta}^f}(\tau)$ are respectively the variance and the normalized autocovariance of returns for $x=0$ and $\sigma_c^2$ and $\rho_{r_{\Delta}^c}(\tau)$ are the same quantities for $x=1$.\\
In the same way the latter is
\begin{align}
\rho_{r_{\Delta}^2}(\tau|x) = \bigg[1+\frac{(1-x)^4}{x^4}\frac{M_f^4}{M_c^4}+6\frac{(1-x^2)}{x^2}\frac{\sigma_c^2\sigma_f^2}{M_c^4}\bigg]^{-1}\rho_{(r_{\Delta}^c)^2}(\tau)+ \nonumber \\
 \bigg[1+\frac{x^4}{(1-x)^4}\frac{M_c^4}{M_f^4}+6\frac{x^2}{(1-x)^2}\frac{\sigma_c^2\sigma_f^2}{M_f^4}\bigg]^{-1}\rho_{(r_{\Delta}^f)^2}(\tau) +\nonumber\\
\frac{2}{3}\bigg[1+\frac{x^2}{6(1-x)^2}\frac{M_c^4}{\sigma_c^2\sigma_f^2}+\frac{(1-x)^2}{6x^2}\frac{M_f^4}{\sigma_c^2\sigma_f^2}\bigg]^{-1}\rho_{r_{\Delta}^c}(\tau)\,\rho_{r_{\Delta}^f}(\tau)\label{volapp}
\end{align}
where $M_f^4$ and $M_c^4$ are the fourth moment of returns in the case $x=0$ and $x=1$.\\
As in the previous section we have to evaluate the following expectation values to elimate the dependence on $x$ in Eq. (\ref{arbapp}) and Eq. (\ref{volapp})
\begin{align}
\rho_{r_{\Delta}}(\tau) = \int_0^1 dx\,\rho_{r_{\Delta}}(\tau|x) f_{eq}(x)\\
\rho_{r_{\Delta}^2}(\tau) = \int_0^1 dx\,\rho_{r_{\Delta}^2}(\tau|x) f_{eq}(x).
\end{align}
Fig. \ref{approxarb} shows the comparison of these two autocovariance functions calculated from a simulation with the corresponding formulas just found. The superposition approximation works quite well for the autocorrelation of the returns but it fails to reproduce the correlations of the squared returns. For the simple returns we conjecture that this agreement is due to the linear balancing of the positive inefficiency of chartists and the negative one of fundamentalists and corresponds really to the incoherent superposition of opposite effects. For volatility clustering instead the simple superposition dose not capture at all the nature of the phenomemon. The coherent persistence of a chartist bubble, in fact, cannot be reproduced by the simple superposition approximation. On the other hand volatility clustering are not present in any of the two limitating cases. \\
\clearpage
\section{Random walk of fundamental price and effective gaussian transition}
\begin{figure}[htbp]
\begin{center}
\includegraphics[scale=0.35]{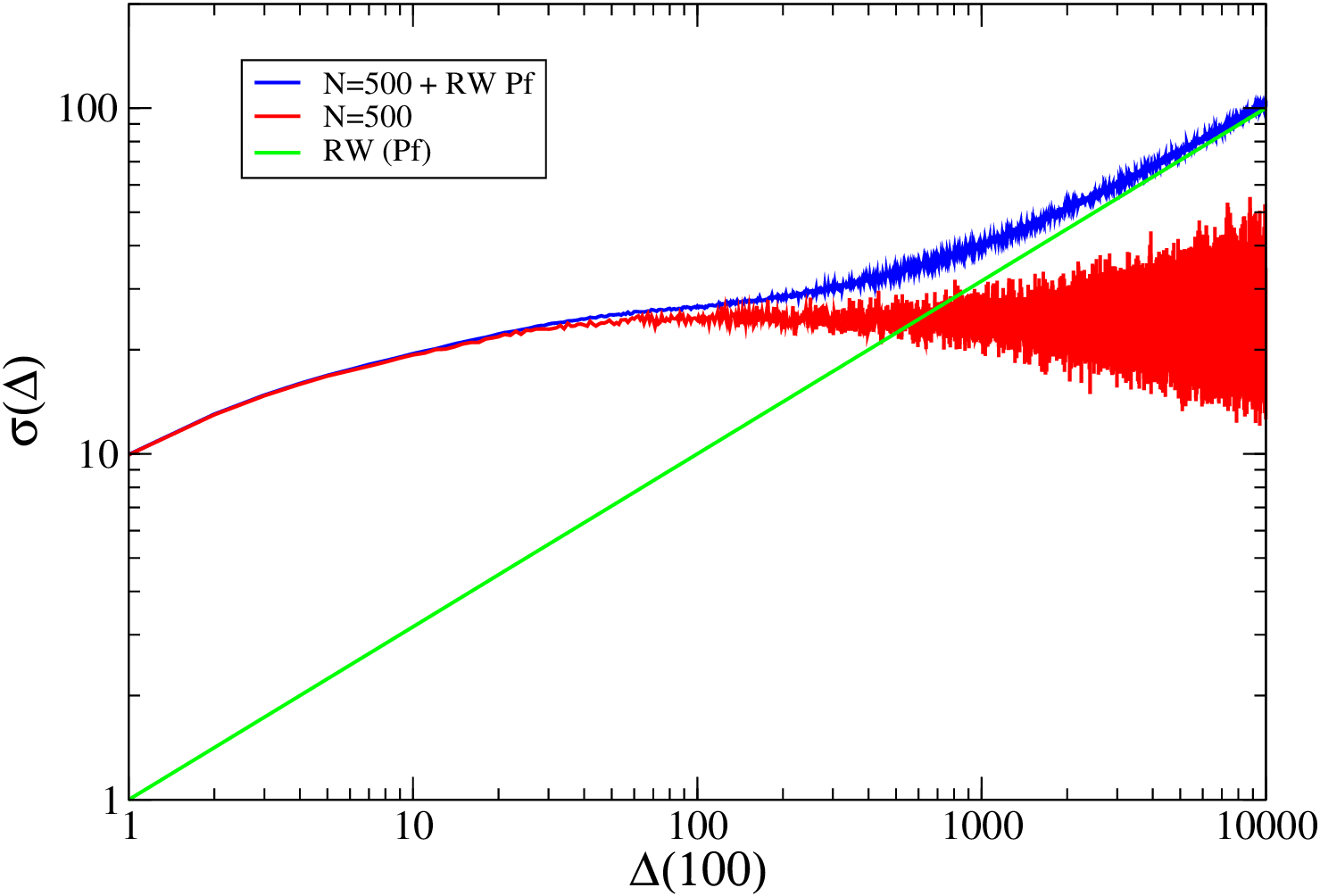}
\caption{We show the same analysis of Fig. \ref{diffulong} for the scaling behavior of the variance of the returns in the case in which the fundamental price is not constant (as in Fig. {diffulong}) but it follows an intrinsic random walk  with $\sigma_{pf}=1/10$). We note, as expected, that when $\Delta\rightarrow\infty$ the leading contribution comes from the random walk of $p_f$. Therefore in our interpretation the transition to gaussianity of real data at long times arises from the predominance of the $p_f$ random walk with respect to the Stylesd Facts which become extremely rare events at long times.}
\label{diffu_rwpf}
\end{center}
\end{figure}

\begin{figure}[htbp]
\begin{center}
\includegraphics[scale=0.35]{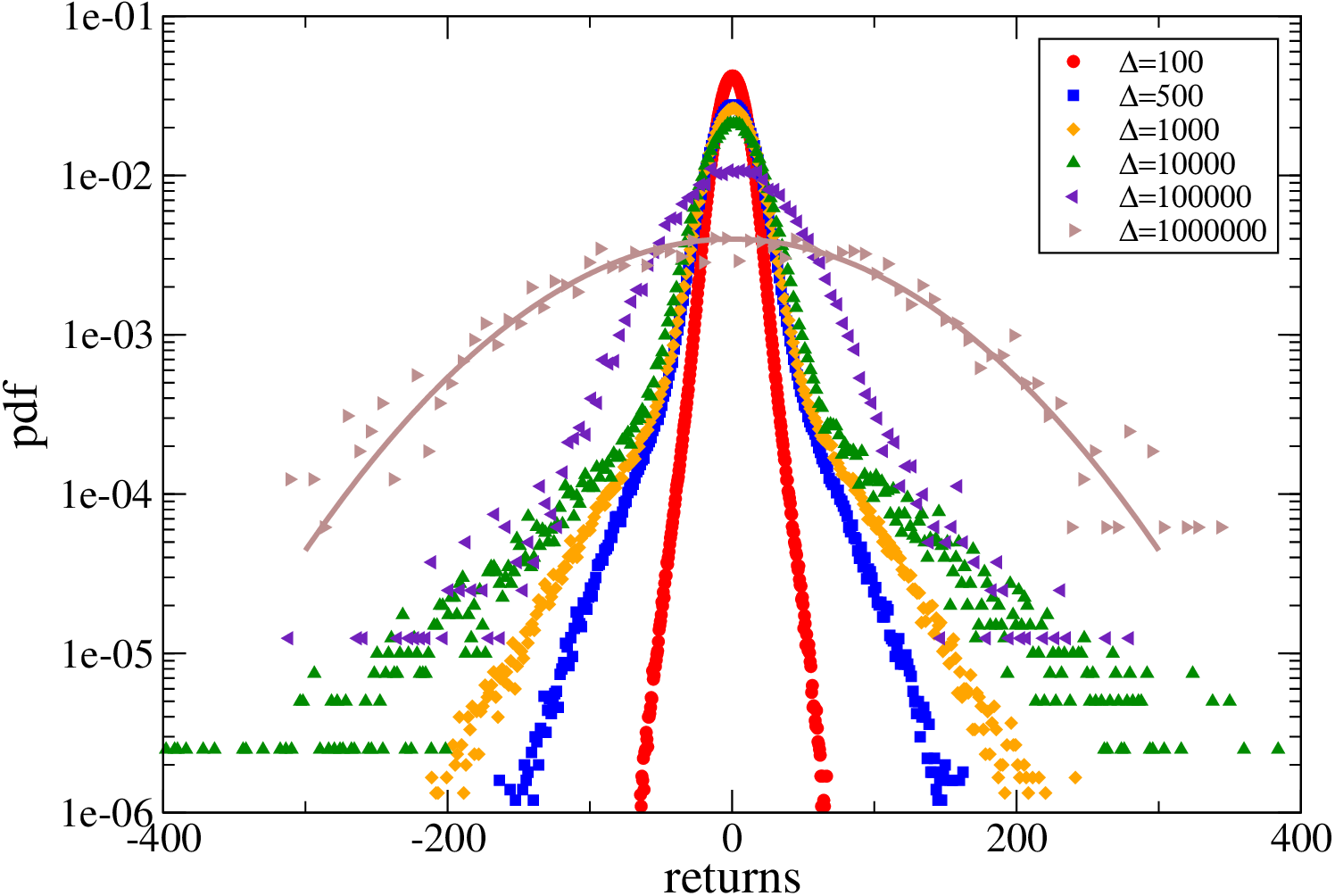}
\caption{The panel shows the comparison of the return probability density function for different $\Delta$. When $\Delta =10^6$ this function is well fitted by a gaussian (solid line). The effective transition to the gaussian behavior derives from the random walk of $p_f(t)$.}
\label{allpdf_rwpf}
\end{center}
\end{figure}

\begin{figure}[htbp]
\begin{center}
\includegraphics[scale=0.4]{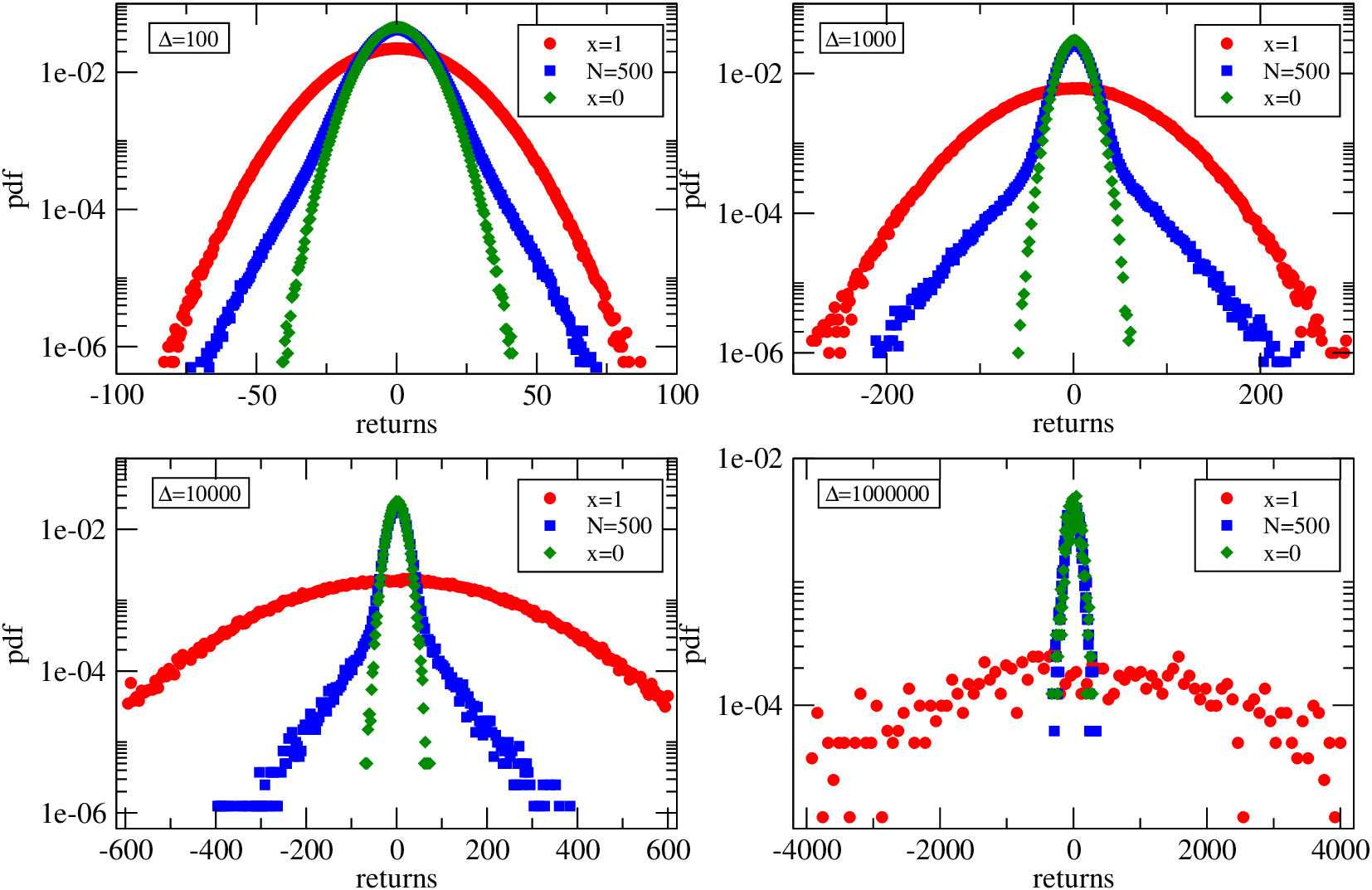}
\caption{The random walk of $p_f$ makes the return probability density function of the case $x=1$ wider and wider therefore, when $N=500$, the deviation from the density of the case $x=0$ is not observed.}
\label{pdfCFN_rwpf}
\end{center}
\end{figure}

We have seen that our minimal model, strictly speaking, does not lead to gaussian transition for the returns at long times. The fact that real data seem to show such a transition  therefore may appear as problematic. On the other hand the model was aimed at the identification of the nature of the SF and we made the simple assumption that $p_f=\text{cost}$. Clearly this is unrealistic and for a real comparison with the data one should consider also the dynamics of $p_f(t)$. This will change the situation in an important way and will reconcile this apparent discrepancy. \\
Following this reasoning we could expect that the return pdf gets very similar to the pdf of the process of $p_f(t)$ when the term due to the drift becomes leading. Hence a natural choice for the $p_f$ consists in a random walk with a variance smaller than the variance of the white noise of Eq. ($\ref{price2}$). How much it is smaller just fixes the time scale $\Delta^*$ at which the drift becomes the leading term. \\
In Fig. \ref{diffu_rwpf} we report the same diffusional analysis of Fig. \ref{diffulong} in the case of $\sigma_{p_f}=10^{-1}\sigma$ where $\sigma$ is the same parameter of Eq. \ref{price2} and Fig. \ref{allpdf_rwpf} shows the pdf for various value of $\Delta$. The main difference with Fig. \ref{allpdf1} is represented by the brown square curve which is well-fitted by a gaussian without appreciable tails. This results is obvious because we have choosen that $p_f$ makes a random walk and the brown curve is in the regime $\Delta>\Delta^*$. The random walk of $p_f$ makes the variance a strictly increasing function of $\Delta$ (for $\Delta>\Delta^*$) and the data in Fig. \ref{pdfCFN_rwpf} provide a clear evidence that the dissappearence of the non gaussian tails is due to a statistically irrelevance of them when an additional gaussian term is present. In this respect one cannot really say that Fat Tails disappear but they become extremely rare events and are essentially obscured by the gaussian behavior of $p_f$.
\clearpage
\section{Multiplicative version}\label{moltip}
The role of Walras'  law (see \cite{kreps}) in economics is very similar to the role of Newton's law of the dynamics in physics because Walras' law introduces a relationship between the excess demand (ED) and the consequent variation of the price that ED causes (the excess demand is usually defined as a function of the unbalance of demand and supply). This law has been formalized in many ways since it was introduced about two centuries ago and the version, which belongs to the framework of our model, is the following
\begin{equation}\label{walras1}
\frac{1}{p}\frac{dp}{dt}=\beta\,ED
\end{equation}
or when time is discrete ($\Delta t=1$):
\begin{equation}\label{walras2}
\frac{\Delta p}{p}  =\beta\, ED
\end{equation}
The basic idea of Eq.  (\ref{walras1}) is that price follows a multiplicative process rather than a linear one so its percentage increments, rather than the increments themselves, are proportional to ED. There is appreciable evidences in the experimental data that a multiplicative dynamics is indeed more appropriate than a linear one\cite{mantegnabook,bouchaud}.\\
Up to now we have neglected this multiplicative nature in our model and we have adopted the simplified linear dynamics. As we are going to see in this section the two cases essentially coincide for small fluctuations. The purpose of our model was to trace back the detailed origin of the SF and their self-organization. In this perspective the linear dynamics does not change the essence of these elements but it represents an useful simplication especially in deriving some analitical results. In this section we discuss in detail the analogies and differences of the two cases.\\
In our model the (linear) price increments are proportional to the term $- x\,b/(M-1)(p_t-p_M)+(1- x )\gamma(p_f-p_t)$ which we identify as the ED, in this respect it corresponds to a linearized Walras' Law. Considering this as the small price increments limit of a multiplicative dynamics we can now try to go back and consider the real dynamics from which this approximation may arise. We are going to see that this backward reconstruction from linear to multiplicative presents some subtle points which are worth to discuss in some details.\footnote{in section 7 and in all its subsections returns are defined as $r_{t+\Delta}=\ln(p_{t+\Delta}/p_t)$}
\subsection{Choice of excess demand}
\begin{figure}[htbp]
\begin{center}
\includegraphics[scale=0.35]{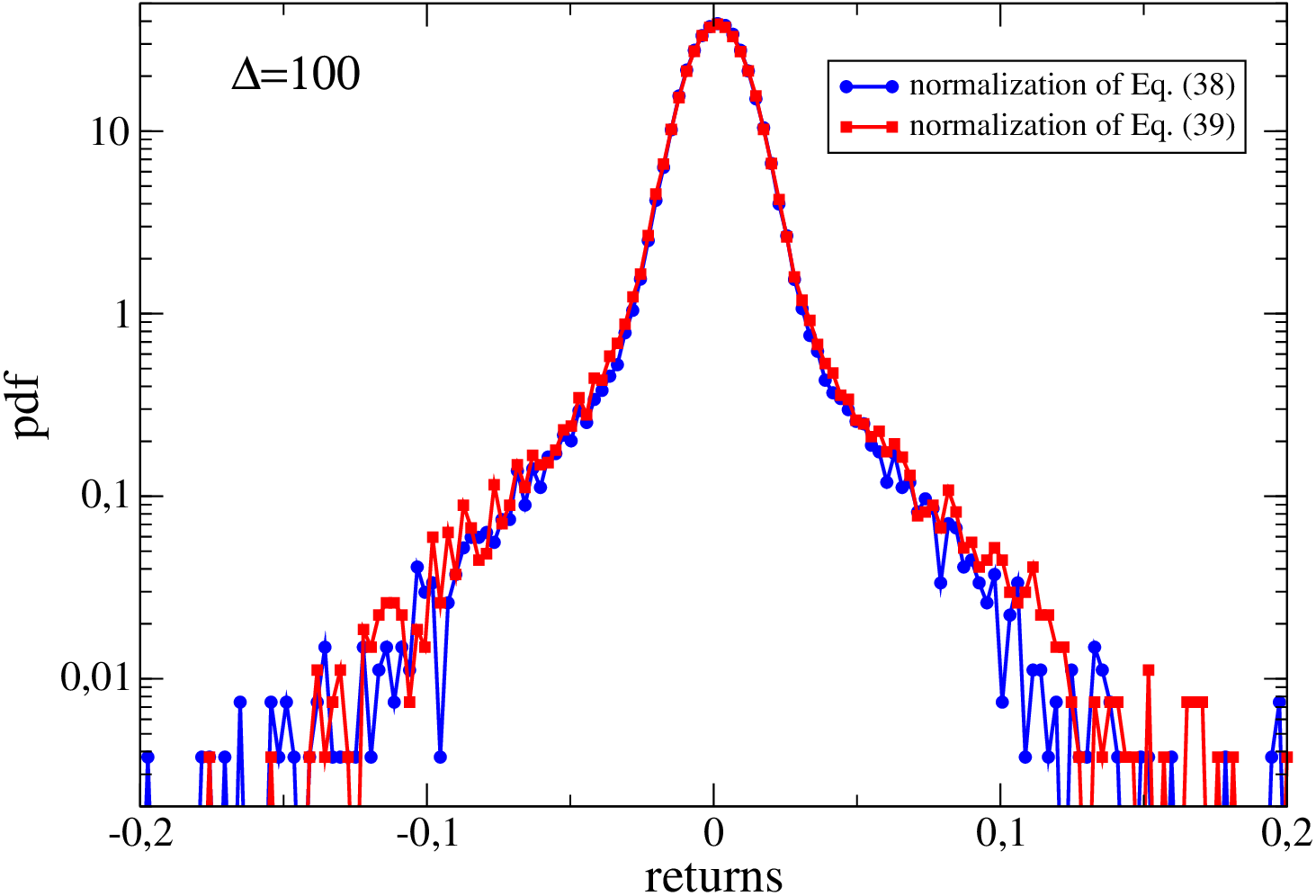}
\begin{verbatim}



\end{verbatim}
\includegraphics[scale=0.35]{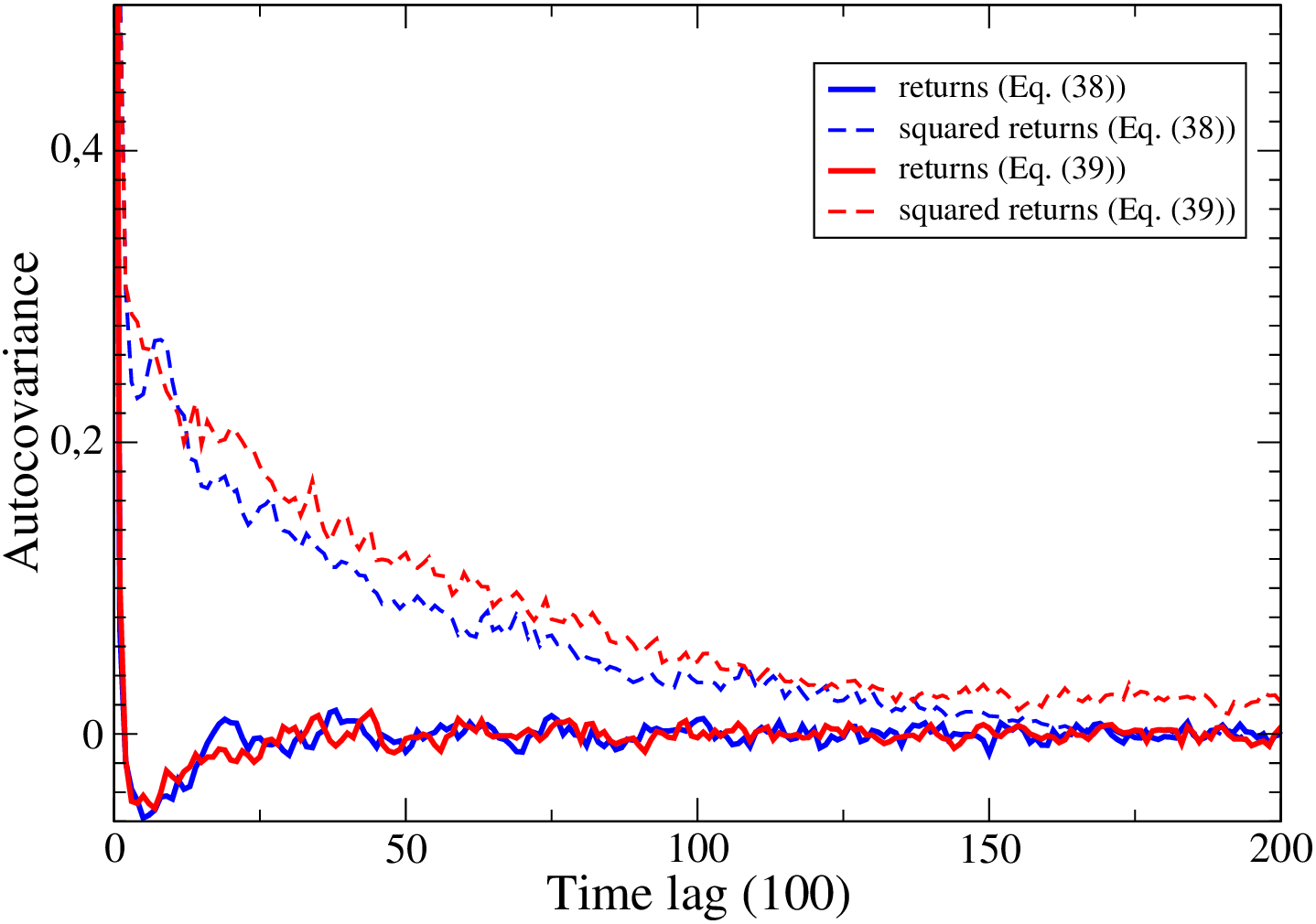}
\caption{The two different normalization for ED lead to similar Stylised Facts. In our opinion (see text) the most natural choice is to normalize ED with the price itself which is an objective piece of information shared by all agents.}
\label{cfr_norma} 
\end{center}
\end{figure}

We would like now to construct a price equation which is consistent with Eq. (\ref{walras2}) so we define the price increments as a geometric random walk with an additional term that is indeed the excess demand:
\begin{equation}\label{multi}
\ln(p_{t+1})=\ln(p_t) + ED + \sigma \xi_t
\end{equation}
where $\xi_i$ are independent  and normally distributed random variables. However Walras'  law does not specify which is the correct choice for ED, apart from the fact that the dimensional analysis of Eq. (\ref{walras2}) implies that ED must be somehow normalized (dimensionless).
The term of Eq. (\ref{price2}) that plays the role of ED  has the dimension of a price so the simpliest way to make the ED of the linear version dimensionless is to normalize it with the quantity which has the dimension of a price. This normalization is not univocous and for example one could have the two different situations
\begin{align}
ED= x  \frac{b}{M-1}\frac{(p_{t}-p_M)}{p_t}+ (1- x )\gamma\frac{(p_f-p_t)}{p_t} \label{norma1}\\
ED= x  \frac{b}{M-1}\frac{(p_{t}-p_M)}{p_M}+ (1- x )\gamma\frac{(p_f-p_t)}{p_f} \label{norma2}
\end{align}
with obviously $p_f \ne 0$.\\
In Eq. (\ref{norma1}) chartists and fundamentalists have the same benchmark to calculate the percentage ED while in Eq. (\ref{norma2}) each of them has different one: chartists use the moving average, fundamentalists use the fundamental price. In Fig. \ref{cfr_norma} we show and compare the results in the two cases. The behavior is very similar and the detectable feature is an increase of the fluctuations in the case of Eq. (\ref{norma2}). \\ 
\\ 
Despite these similarities the two normalizations correspond to a qualitatively different nature of the dynamics. In fact suppose, for example, that the price has just reached the local minimum. Then it begins to grow and so it crosses the moving average from below. Once crossed, the following inequality holds $(p_t-p_M)/p_M > (p_t-p_M)/p_t$. Conversely if the price has just reached the local maximum the opposite inequality is $(p_t-p_M)/p_M < (p_t-p_M)/p_t$. This situation would imply that in the limit dominated by the chartists the price will unavoidably go to infinity for Eq. (\ref{norma2}) or to zero for Eq. (\ref{norma1}) depending on the initial condition. This simple example shows that the two normalizations can affect in a very different way the large fluctuations of the dynamics. In our opinion the most appropriate normalization is the one corresponding to Eq. (\ref{norma1}) because $p_t$ is a natural value recognized by all agents and in addition this leads to the correct linearized dynamics as now we are going to see.
\subsection{Calibration of parameters and model instabilities}
\begin{figure}[!t]
\begin{center}
\includegraphics[scale=0.4]{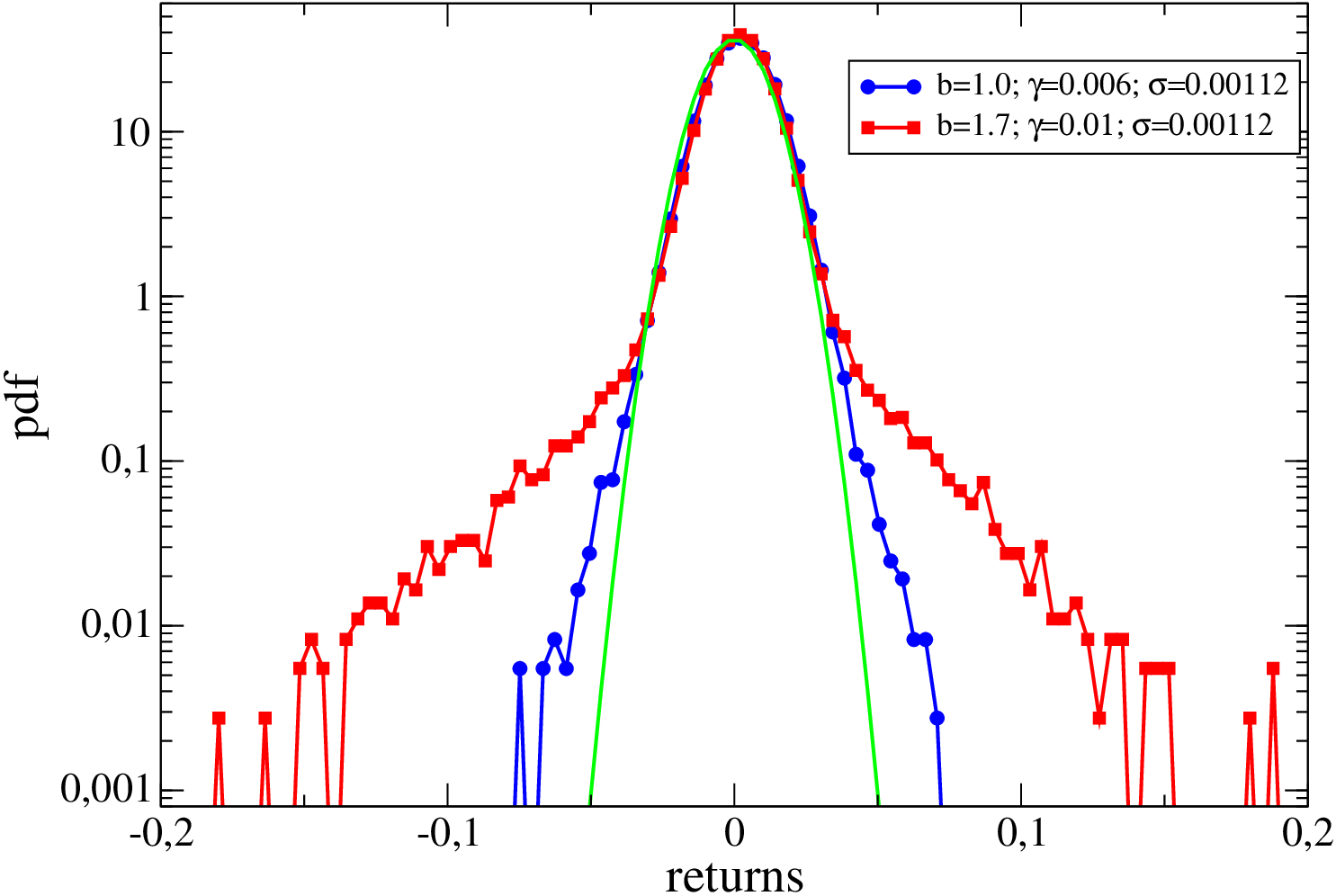}
\caption{The multiplicative process results appear to be extremely sensitive of the region of parameters choosen. Here we show two slightly different set of parameters which lead to drastically different results with respect to the Fat Tails phenomenology.}
\label{2regimes}
\end{center}
\end{figure}

Most of the ABMs share the common aspect that they (more or less) reproduce the empirical facts (i.e. SF) only in a certain region of the phase space of their parameters.  A tipical example is the model of Lux and Marchesi \cite{Lux:1999,LM} or \cite{paperoI,paperNP}. Therefore they require a fine tuning of all the parameters and even of the total number of agents $N$ in order to reproduce the SF.\\This situation is already present in the models with the linear dynamics. The multiplicative version, as for example in our model Eq. (\ref{multi}), is dramatically unstable and more sensitive to small changes of parameters than the linear one. The results is that the dynamics of price cannot be simulated for a large part of the phase space of the parameters because it is subject to extreme singularities towards zero or infinity. Consequently the main attempt of tuning the parameters consists in finding the subtle equilibrium between the weight of the noise (its variance $\sigma$) and the weight of the ED ($b$ and $\gamma$) in the evolutive Eq. (\ref{multi}) of the price.\\
\\
The starting point of this non trivial calibration is the choice of a reasonable value of the variance of the white noise. In fact if the variance $\sigma$ would be of order $1$ the typical increment in a single time step would be about 100\% of the price or more, which is totally unreasonable. On the other hand, in order to calibrate the noise we should relate the model time step to real time. The time scale of our model is expressed in arbitrary units and these should be tuned with respect to the realistic dynamics. If we would like to intepret our model time step as a single elementary operation in the real market this can be done in the following way. The typical price fluctuations of the price of a stock during one day is of the order of 1-2 \%. The number of individual operations for a typical stock is of the order of $1000 - 5000$. From these heuristic arguments one can the argue that a typical variance in the model should be of the order $\sigma=0.001$.\\
Then we should set the parameters $b$ and $\gamma$ so that the magnitude of ED is comparable with the white noise. In fact if it would happen $<ED^2>\ll \sigma^2$ the process will be dominated by the gaussian noise and it would not show the non gaussian tails. In the other hand if the weight of ED is too large we will observe the intrinsic unstability of the process due to the exponential dependence on ED in Eq. (\ref{multi}). We find that the upper values that can be choosen are $b\approx1.7$ and $\gamma\approx0.01$. In Fig. \ref{2regimes} we compare the tails of the return pdf for two set of parameters $\{b=1.7, \gamma=0.01, \sigma=0.00112\}$ and  $\{b=1.0, \gamma=0.006, \sigma=0.00112\}$. It is important to remark that the former set gives a significative deviation from the gaussianity in the sense of the SF, while the latter set instead produces a pdf which is very close the original gaussian. \\
We have argued that the linear version should be a linearization of the original Walras' law Eq. (\ref{walras2}). If we now linearize Eq. (\ref{multi}) with the normalization of Eq. (\ref{norma1}), it follows that
\begin{equation}\label{lineare}
p_{t+1}-p_t \approx (\sigma p_t) \xi_t +   x  \frac{b}{M-1}(p_{t}-p_M)+ (1- x )\gamma(p_f-p_t). 
\end{equation}
This linearization is justified if $(p_{t+1}-p_t)/p_t \ll 1$ and this condition is generally true for our choice of parameters. The comparison of Eq. (\ref{lineare}) with Eq. (\ref{price2}) highlights that the only difference lies in the variance of the white noise. This variance is a random variable in Eq. (\ref{lineare}) and the linear version Eq. (\ref{price2}) is essentially the linearization of the multiplicative dynamics if the variance of the white noise can be assumed as nearly constant and its stochastic nature negletcted. Let us define the effective variance as $\hat{\sigma}=\sigma p_t$, the random variables $\xi_t$ and $p_t$ are independent beacuse $p_t=p_t(\xi_0, \xi_1,\ldots,\xi_{t-1})$, hence we can focus just on the effective variance $\hat{\sigma}$. In the average $E[\hat{\sigma}]=p_f\sigma$ and the typical fluctuation $\Omega$, normalized with the squared average, is
\begin{equation}\label{varianzamulti}
\Omega = \sqrt{\frac{E[p_t^2]}{p_f^2}-1}\approx \sqrt{\frac{E[p^2]}{p_f^2}-1}
\end{equation}
The linear process of Eq. (\ref{price2}) is therefore justified if the fluctuations given by Eq. (\ref{varianzamulti}) are negligible. In other words Eq. (\ref{price2}) is the limit of the linearized Eq. (\ref{lineare}) when $\Omega$ goes to zero. In table. \ref{tabomega} we show that fluctuations around the mean value of $\hat{\sigma}$ are very small independently on $p_f$.
\begin{table}[htdp]
\begin{center}
\begin{tabular}{|c|c|}
\hline
$p_f$ &$\Omega$ \\
\hline
1 &0.029\\
10& 0.031\\
50& 0.032\\
100& 0.034\\
500& 0.034\\
1000& 0.035\\
\hline
\end{tabular}
\end{center}
\caption{Estimation of $\Omega$ for different value of $p_f$.}
\label{tabomega}
\end{table}
This shows that for a choice of the parameters which appears reasonable to reproduce real markets this effect is only of the order of $3\%$. Therefore the linearized version of the dynamics indeed is a valid approximation to the multiplicative one. It is important to note however that a different choice of the parameters could lead to a situation in which even in the limit of small price increments the simple linearized (with the random noise) form would be incorrect. 
Unfortunately we are not able to perform an analitical approach for the multiplicative version as we have done in section 2 because of two reasons: the first one is the non linearity of Eq. (\ref{multi}) which makes the problem analitically very difficult, the second one is that the process cannot be simulated for $ x =1$ because the price goes quickly to zero or grows indefinitely according to the normalization used.

\subsection{Stylized Facts}
\begin{figure}[!htbp]
\begin{center}
\includegraphics[scale=0.38]{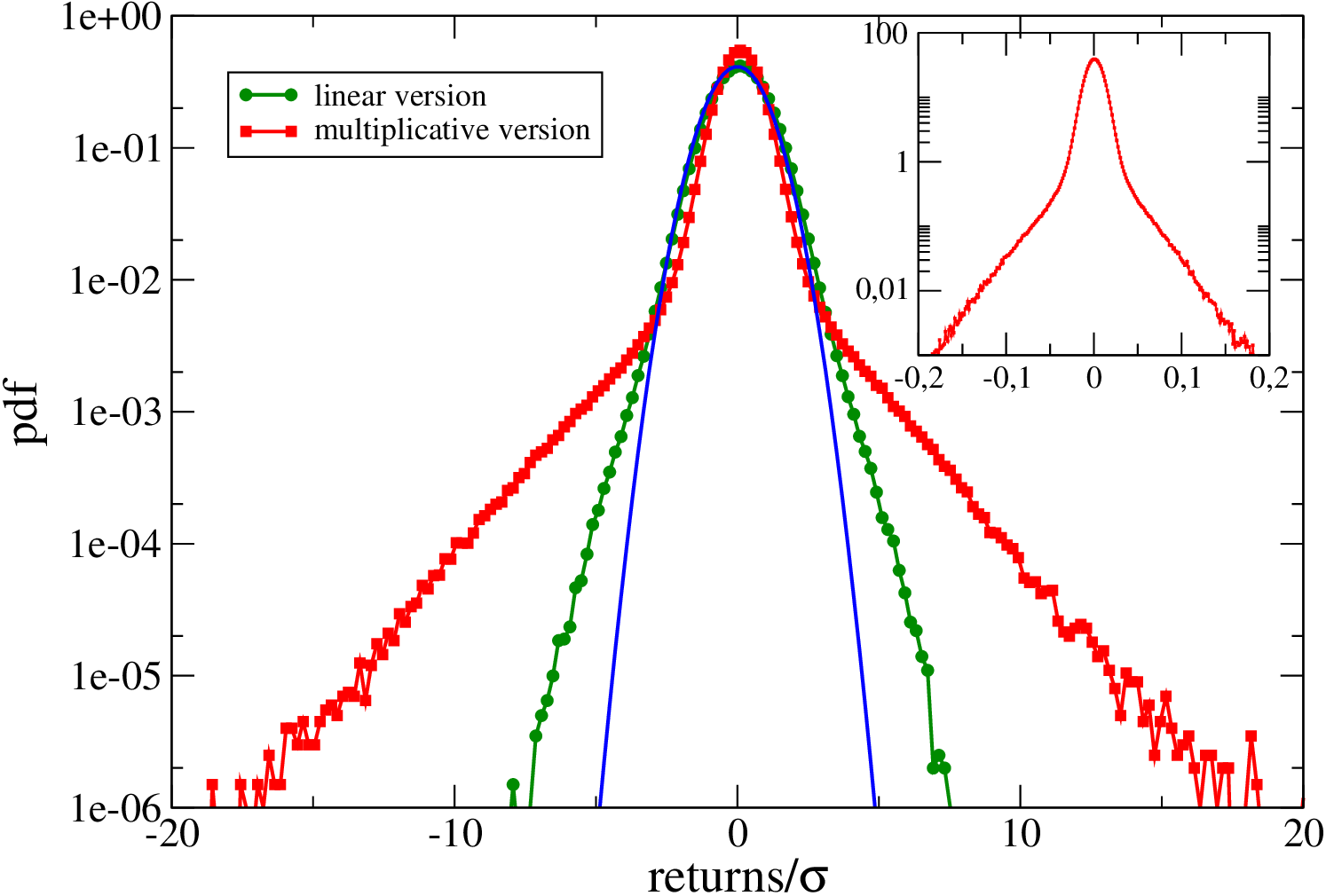}
\caption{Fat Tails for linear and multiplicative dynamics (time interval equals 100 steps). The multiplicative case is characterized by much stronger  non gaussian tails (the returns are normalized with their variance). The solid line represents the gaussian benchmark. The small panel on the right shows the non normalized returns probability density function in the multiplicative case.}
\label{cfr_returns}
\end{center}
\end{figure}

\begin{figure}[htbp]
\begin{center}
\includegraphics[scale=0.38]{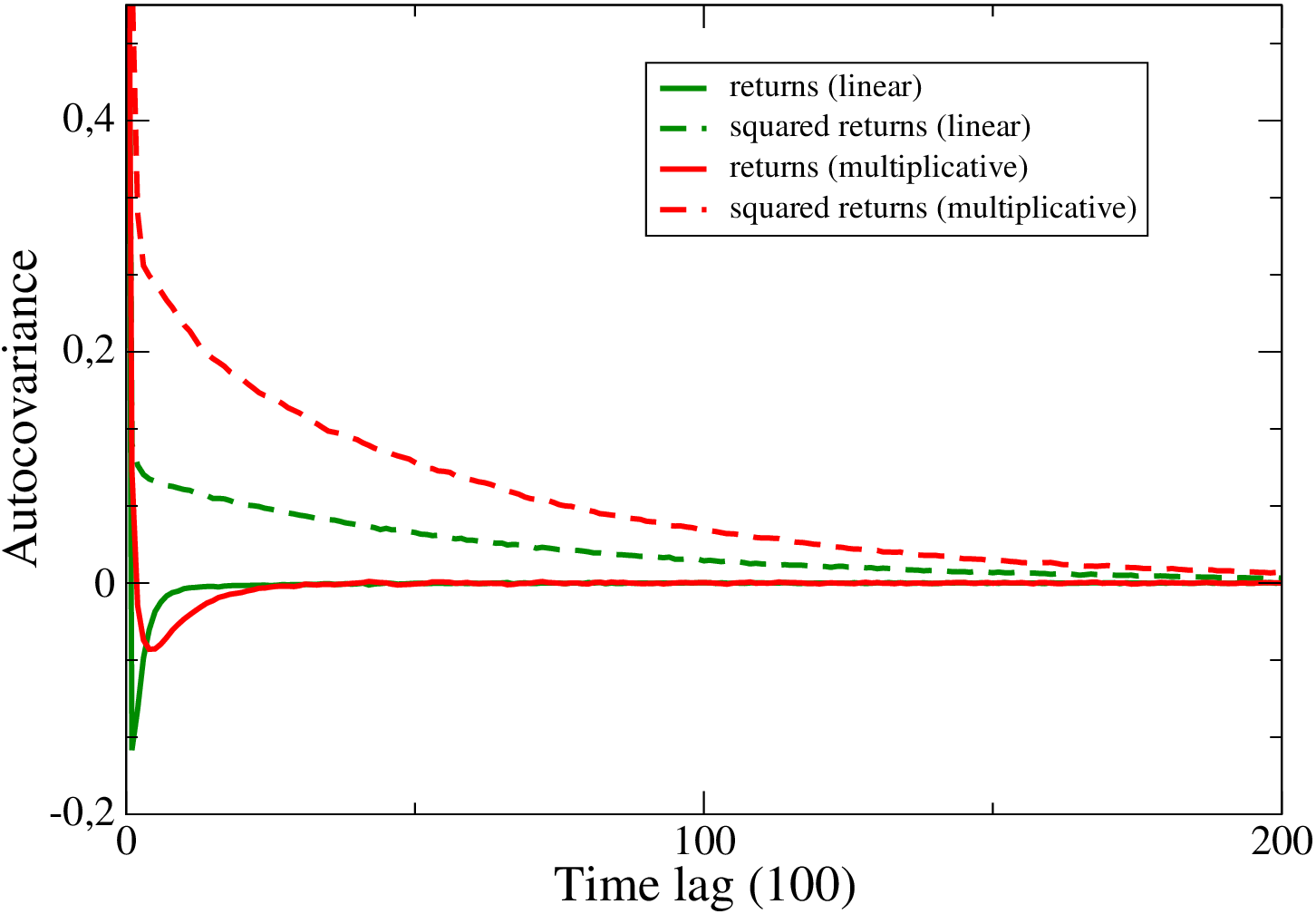}
\caption{Volatility clustering and market efficiency for linear and multiplicative dynamics. The autocovariance function of returns and of squared returns are qualitatively similar in the two cases. The volatility clustering is larger in the multiplicative case than in the linear one. The linear correlation shows very similar behavior. Note that in this minimal version of the ABM both the linear and multiplicative dynamics lead to a single characteristic time scale and therefore volatility clustering has an exponential decay. The possibilities of multiple time scales for the agents can modify this behavior in a more realistic one (power law like) as shown in Fig. {7} of \cite{paperoI}}
\label{cfr_corr}
\end{center}
\end{figure}

The previous discussion may suggest that the multiplicative version introduces only disadvantages because it is much more unstable in comparison with the linear version with respect of phase space of the parameters. The limit $ x =1$ does not exist and even the limit $ x =0$ appears impossible to be studied analitically. In general very few results can be derived analitically for the complete model. But we have also important advantages, the first one is that  the multiplicative dynamics is more realistic. We are going to see in fact that the SF of the multiplicative version are substantially more pronounced with respect to those of the linear version. In particular the multiplicative dynamics shows an interesting relation between the power of returns and their degree of correlation, which is in better agreement with real data. The specific results we are going to show corresponds to the normalization of Eq. (\ref{norma1}). Even though stationarity is certainly weaker than in the linear dynamics also the multiplicative case appears to converge to a quasi-stationary state but with larger fluctuations.\\
In Fig. \ref{cfr_returns} we show the return pdf for the linear version and for the multiplicative one. The non gaussian tails are much more pronounced in the multiplicative case. In order to compare the two cases we have rescaled the returns normalizing with their variance.\\ 
\\ 
In Fig. \ref{cfr_corr} we compare the normalized autocovariance of returns and squared returns for the two cases. The two models give similar results from a qualitative point of view but the degree of correlation of squared returns is nearly twice in the multiplicative case. Volatility clustering is therefore enhanced by the multiplicative dynamics.
\\ 
\\ 
We can also investigate the degree of correlation of $|r|^\phi$ as a function of $\phi$. It is well-known that the largest correlation in empirical data is observed for $\phi=1$ with respect to higher powers (\cite{Cont:2001}). In Fig. \ref{corr_alpha1} and \ref{corr_alpha2} we report the autocovariance functions for different values of $\phi$. The linear version exhibits a maximum in correlation for $\phi=2$ while for the real data (see insert of Fig. \ref{corr_alpha1}) the correlation is larger for $\phi=1$. In Fig. \ref{corr_alpha2} we can see that the multiplicative version leads to a realistic order of the degree of correlation with the maximum $\phi=1$  (consistent with market data).
\\ 
At the moment we do not have a full understanding of the origin of these different behaviors for the two dynamics. A naif intepretation could be to argue that in a multiplicative model returns are generally smaller than 1, differently from the returns in the linear version. On the other hand that our autocovariance functions are normalized so that they are correctly rescaled to be compared. A simple simulation of the linear version with a set of paramaters so that $|r_t|<1$ for most of the time confirms that this arguments is doubtful because the same anomalous order in correlation functions is observed.
\\ 
\begin{figure}[!t]
\begin{center}
\includegraphics[scale=0.4]{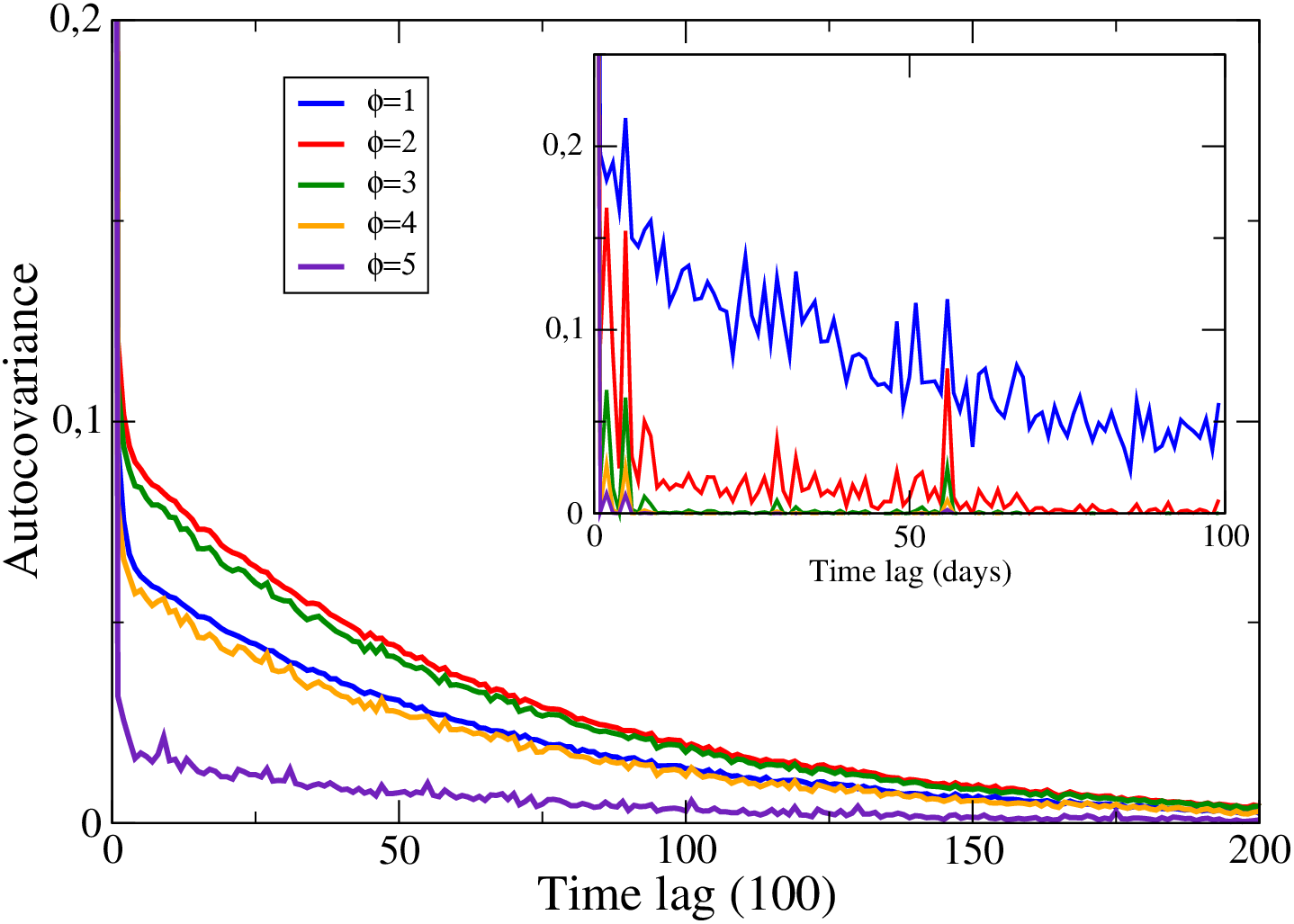}
\caption{Linear model: degree of correlation of absolute returns with respect their power $\phi$. The linear dynamics leads to a maximum correlation for $\phi=2$ while the real data (see insert) show a maximum correlation for $\phi=1$. The real data corresponds to daily dynamics of typical NYSE stocks}
\label{corr_alpha}
\end{center}
\end{figure}
\clearpage
\begin{figure}[!t]
\begin{center}
\includegraphics[scale=0.4]{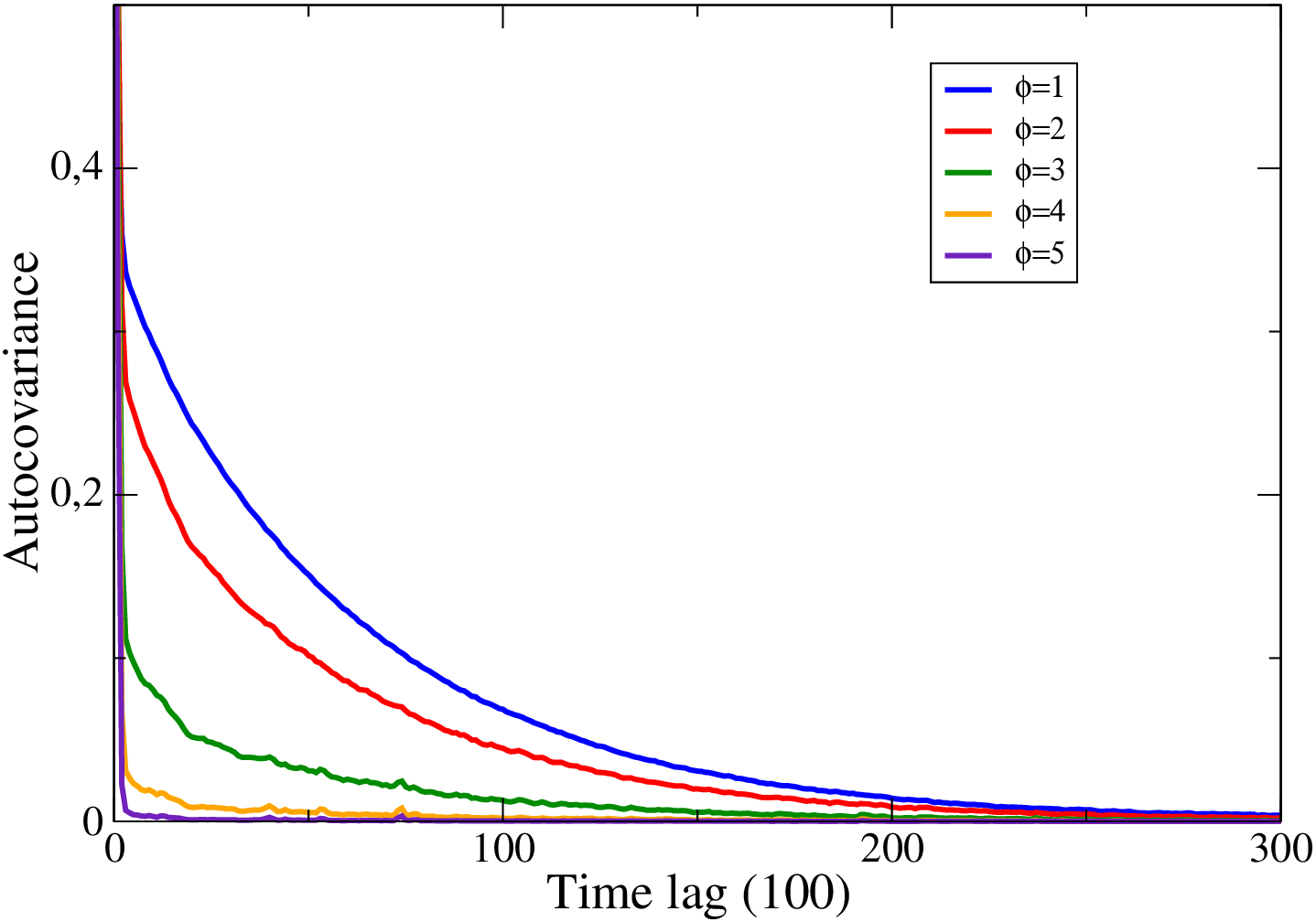}
\caption{Multiplicative dynamics. In this case the order of strength of correlation functions is in agreement with real data corresponding to the maximum for $\phi=1$.}
\label{corr_alpha2}
\end{center}
\end{figure}
\subsection{Diffusional properties and transition to gaussianity}
\begin{figure}[!t]
\begin{center}
\includegraphics[scale=0.4]{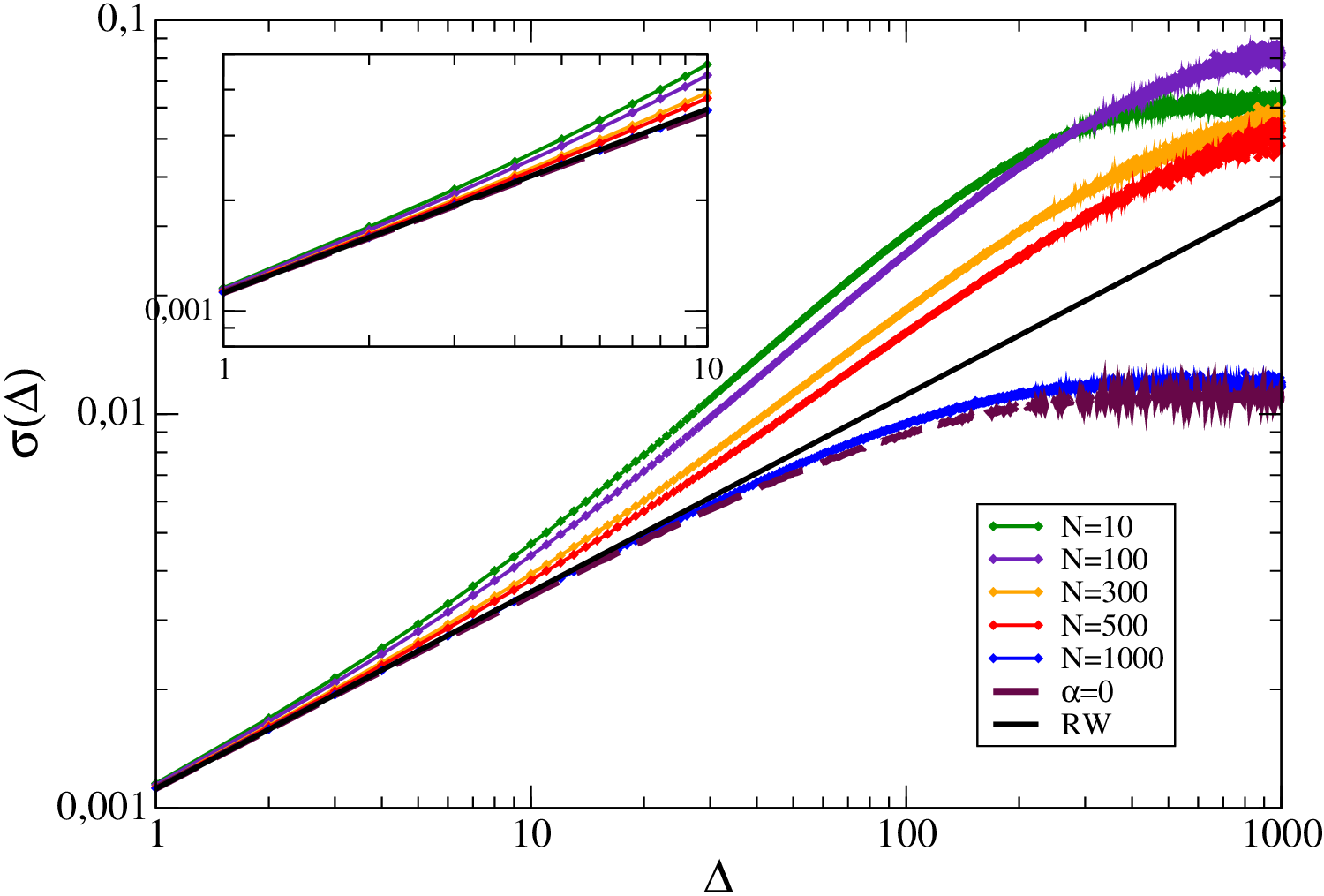}
\caption{The variance of the returns for the multiplicative case is similar to the one observed in the linear case. The diffusion exponent is approximately a decreasing function of $N$.}
\label{diffushortmult}
\end{center}
\end{figure}

\begin{figure}[!t]
\begin{center}
\includegraphics[scale=0.37]{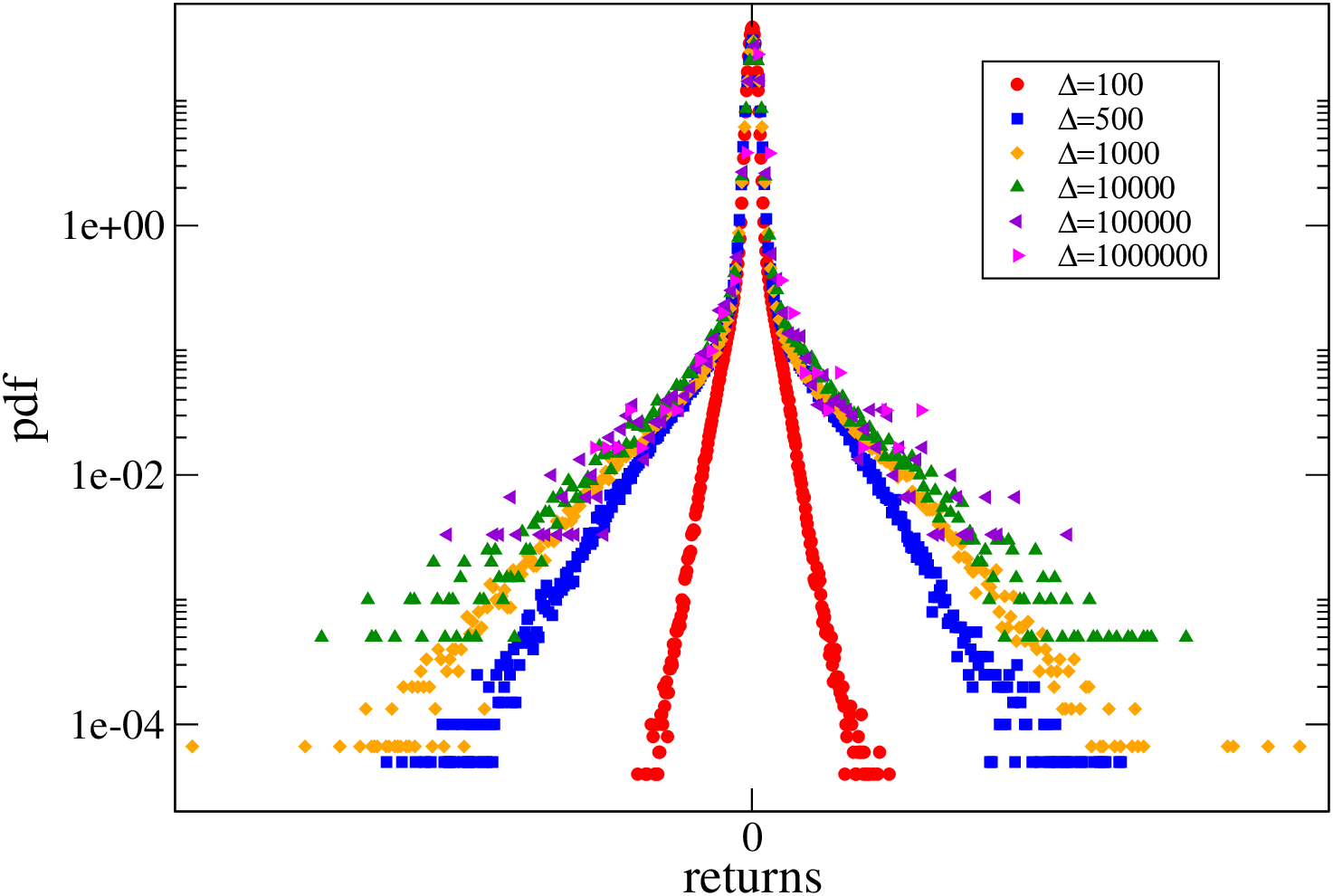}
\caption{Return probability density function for different values of $\Delta$ in the multiplicative case. Once again the tails do not disappear and the transition to gaussianity is not observed. Unless one introduces a gaussian dynamics for $p_f(t)$.}
\label{allpdf1_mult}
\end{center}
\end{figure}

\begin{figure}[htbp]
\begin{center}
\includegraphics[scale=0.37]{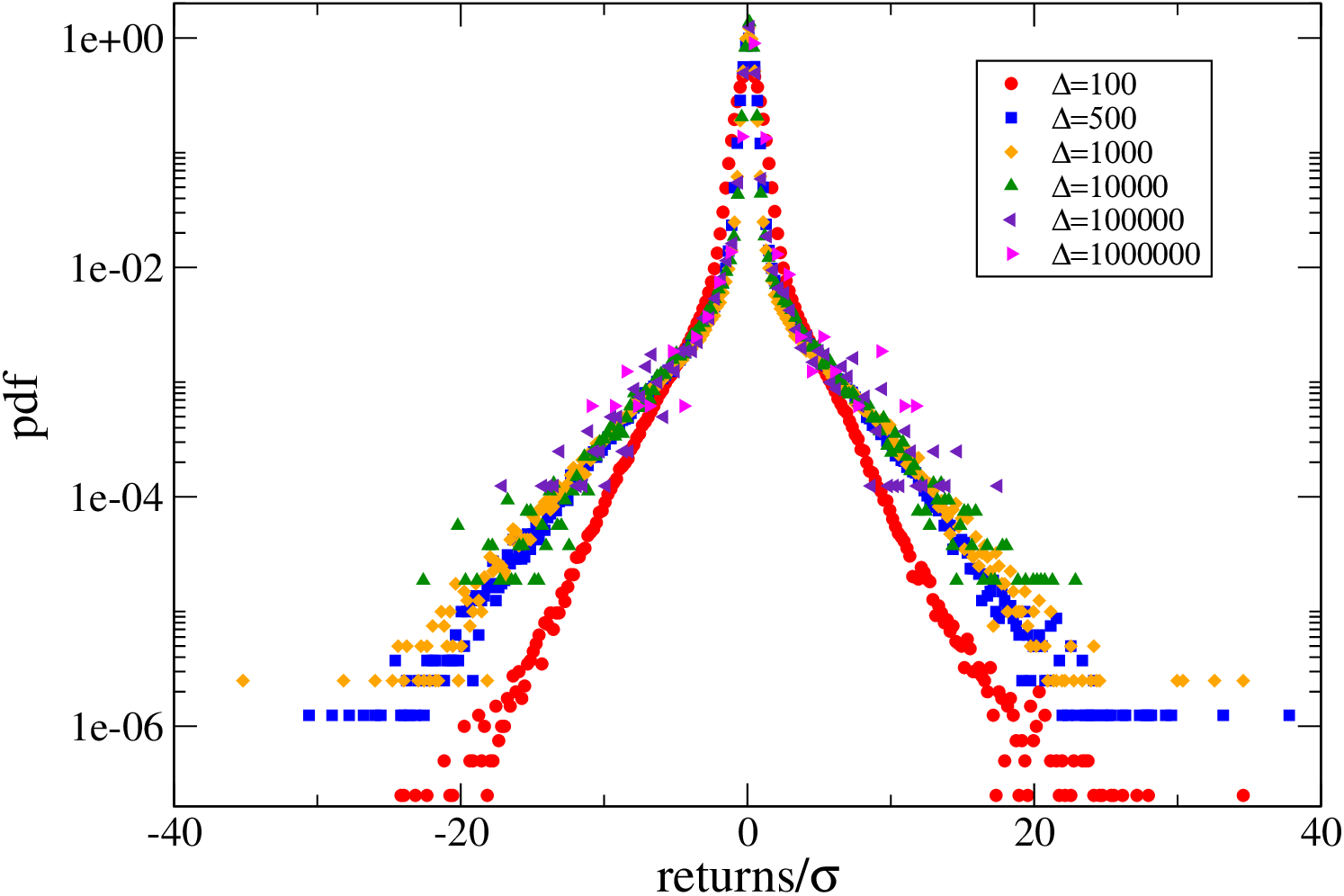}
\caption{The normalized return probability density function shows a better collapse in a universal curve in the multiplicative case with respect to the linear one. However, since in this case the case $x=1$ does not exist we are not able to make a discussion similar to one of Fig. \ref{pdfCFN}.}
\label{allpdf2_mult}
\end{center}
\end{figure}

The second order properties of the multiplicative process and the scaling behavior of the variance of return pdf can be studied in the same way as in section 3 and 4 and all the considerations about the statistical meaning of this analysis are the same except for the fact that the multiplicative process does not allow us to simulate the case $ x =1$ (only chartists). \\
The behavior of the variance of the increments for small values of the time lag $\Delta$ is plotted in Fig. (\ref{diffushortmult}) while table \ref{tabmu_mult} is the analogue of table \ref{tabmu} where we report the results of a fit of the scaling exponent for different N.\\
We obtain the same results of section 3 in which the process has a variance which scales appoximately as a RW for almost any value of $N$.
In the same spirit of Fig. \ref{diffulong} we study the scaling behavior of the variance on a very long temporal horizon and we find that for every $N$ the variance goes to a constant (which depends on $N$) when $\Delta \rightarrow\infty $. The value of this constant values is always larger than the asymptotic value of the case $ x =0$.\\
In a similar way we can also investigate the dependence of the shape of the returns pdf on the time lag $\Delta$. This analysis is reported in Figs. \ref{allpdf1_mult} and\ref{allpdf2_mult}. Once again the multiplicative version confirms our previous observation of persistent non gaussian tails for all values of $\Delta$.\\
The general situation is therefore extremely similar for the multiplicative and linear case. Consequently we expect that also in the multiplicative case the introduction of a RW dynamics for $p_f(t)$ would lead to an effective transition to the gaussian behavior without Fat Tails analagous to the one discussed in section 6. 
\begin{table}[htdp]
\begin{center}
\begin{tabular}{|c|c||c|c|}
\hline
N &$\mu$ &N & $\mu$\\ \hline
 RW & 0.500 &300 & 0.552\\
$ x =0$ & 0.487&400 & 0.535\\ 
$ x =1$ & / &500 & 0.523\\ 
10 & 0.632&1000 & 0.492\\
100 & 0.606&5000 & 0.489\\
250 & 0.561& &\\
\hline
\end{tabular}
\end{center}
\caption{Results of the exponent $\mu$ for $\Delta \in [1,10]$}
\label{tabmu_mult}
\end{table}
\\
\clearpage
\clearpage
\subsection{Self-Organized Intermittency in the multiplicative dynamics}
\begin{figure}[!t]
\begin{center}
\includegraphics[scale=0.35]{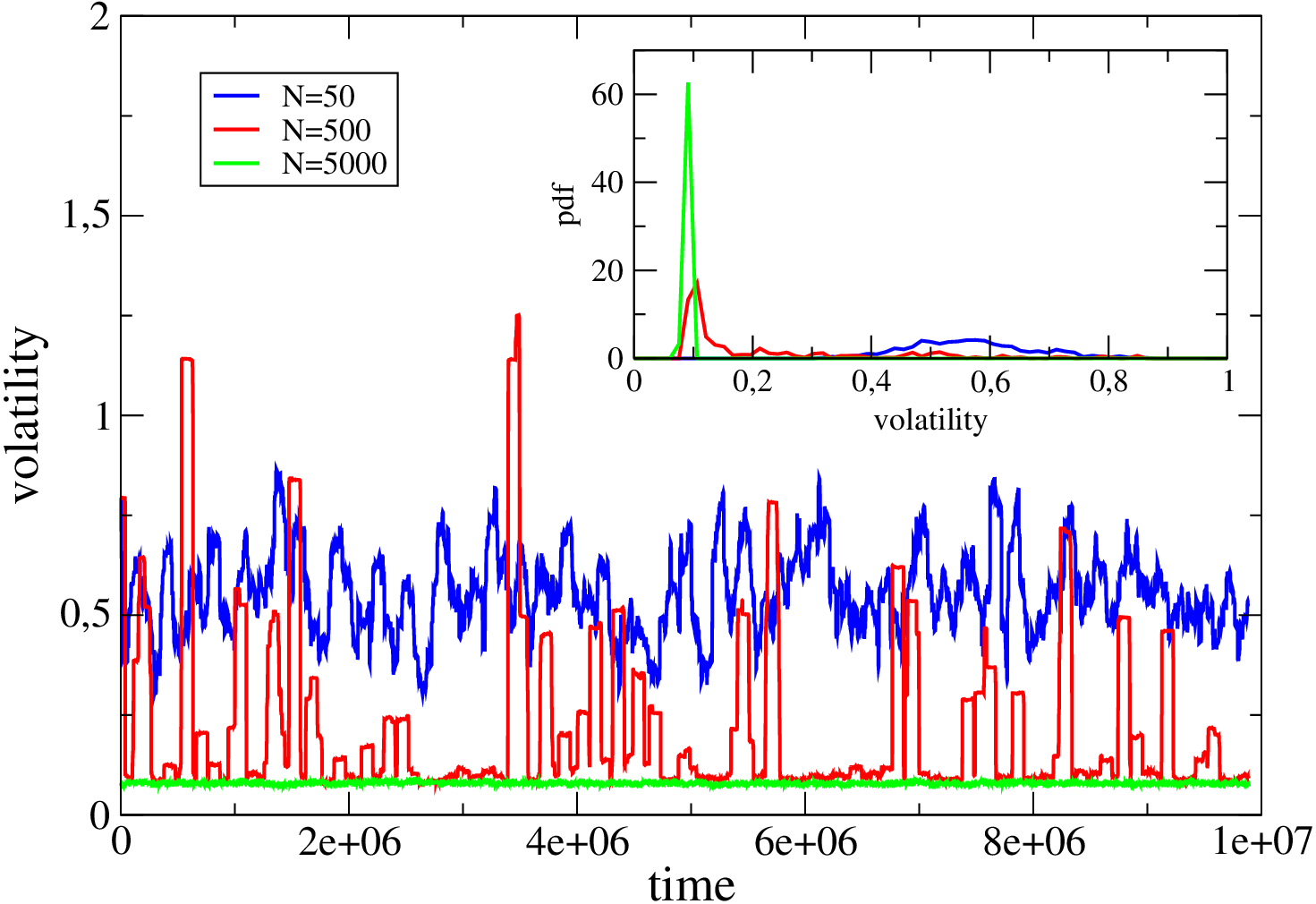}
\caption{Volatility fluctuations for various vlaues of $N$. As in \cite{paperoI, paperNP} we see that for large values of N ($N=5000$), the volatility is very small and this situation is not interesting for the agents. For small values of N (N=50), instead, the volatility is always very high and the market offers arbitrage opportunities for the agents. If $N=500$ we observe an intermittent behavior as in real market. The insert reports the histogram of the main plot.}
\label{ao_mult22}
\end{center}
\end{figure}

Considering that multiplicative case shares the same population dynamics and a very similar behavior with respect to $N$ with the linear version, we can consider the problem of the self-organization along the same lines. 
Thus the key point is again to allow that $N$ can be, in its turn, a variable which is not necessarily fixed a priori. The basic idea is that an agent generally perceives small fluctuations to be not very interesting to trade while large fluctuations can be caught as the sign of trading opportunities. As in \cite{paperNP} we introduce an indicator  $\sigma(t,T)$ of fluctuations which agents consider.
\begin{equation}
\sigma(t,T)=\frac{1}{T-1}\sum_{i=t-T}^{t}(p_i-\bar{p})^2
\end{equation}
The agents enter or exit from the market depending on whether these fluctuations are larger or smaller than some thresholds
\begin{align}
\sigma(t,T)>\Theta_{in}\label{IN}\\
\sigma(t,T)<\Theta_{out}\label{OUT}.
\end{align}
Clearly the thresholds $\Theta_{in}$ and $\Theta_{out}$ are different from those in the linear case and should be suitably adjusted to the properties of the multiplicative dynamics.\\
We report in Figs. \ref{ao_mult22}  the same analysis for the multiplicative case of Fig. 4 of \cite{paperNP} for the linear case. In full analogy with the linear case we can see  that also in the multiplicative dynamics the volatility fluctuations decrease by increasing $N$ and only in an intermediate value of $N$ ($N\approx500$) corresponds to the intermittent dynamics which leads to the SF. This situation, toghethet with the threshold of Eq. \ref{IN} and Eq. \ref{OUT} leads to the self-organization of the system around the quasi- critical value of $N$ ($N^*$), independently on the initual values $N(t=0)$. In fact if we start from a small value of $N$ (i.e. $N\ll N^*$)  the large fluctuations will attract new incoming agents. Conversely if $N\gg N^*$ the typical small fluctuations makes the number of agents decrease. When $N$ reaches $N^*$ the system is relatively stable: the value of $N^*$ depends on the the choice of the parameters, mainly on the thresholds and it correspons to the intermittent state with the SF.\\
The overall picture is very similar to the linear case with the only difference that the convergency to the quasi-critical is somewhat slower with respect to the linear case for large $N$.   
\begin{figure}[htbp]
\begin{center}
\includegraphics[scale=0.35]{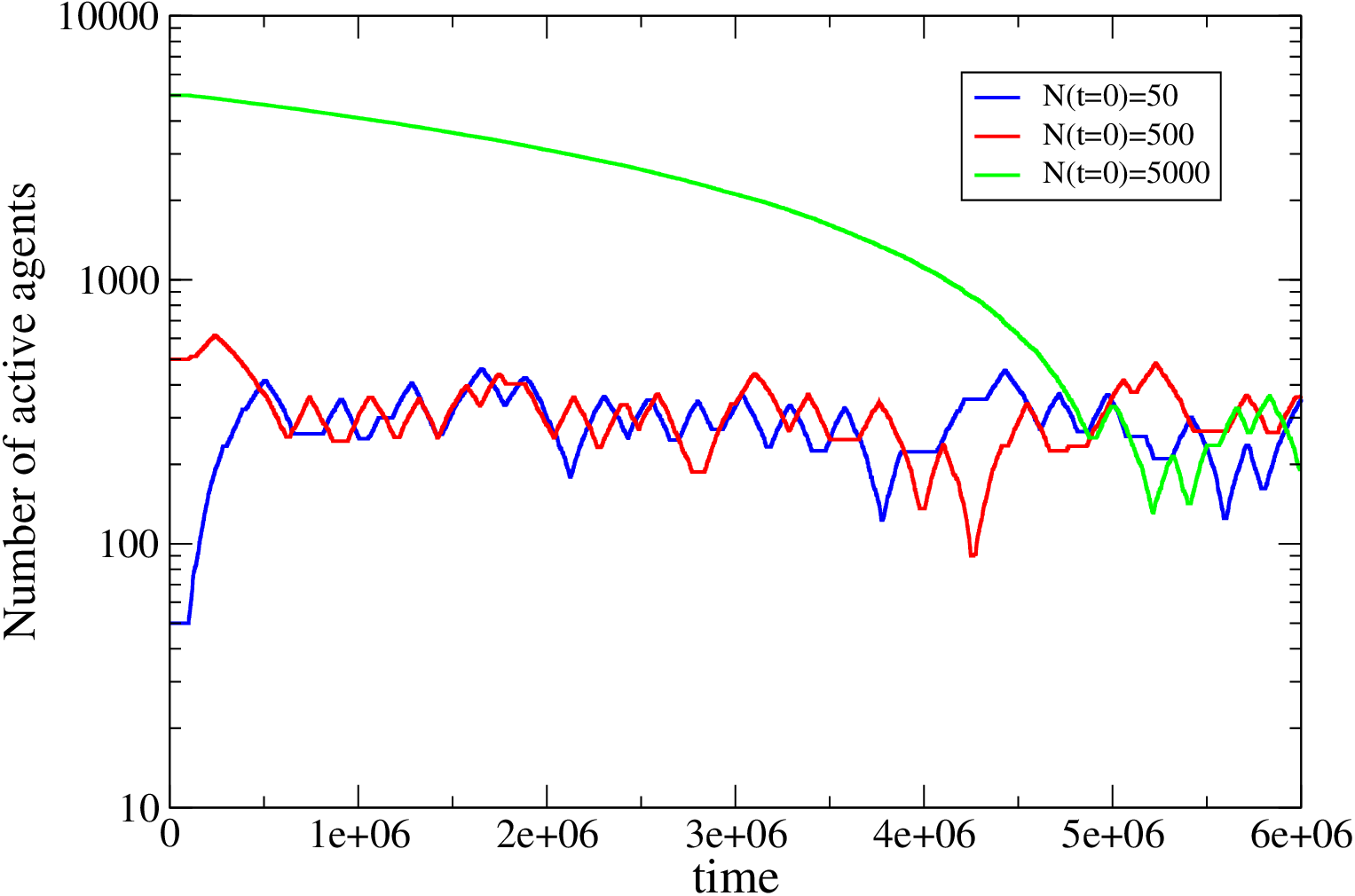}
\caption{Self-Organized Intermittency. The mechanism of self-organization is the same of the linear case proposed in \cite{paperoI,paperNP}. When $N$ is larger than $N^*$ ($N^*\approx 500$ is the number of agents that shows SF) fluctuations are typically lower than $\Theta_{in}$ therefore agents leave the market. Instead when $N$ is smaller than $N^*$ the fluctuations are larger than $\Theta_{out}$ and the agents enter into the market.}
\label{ao_mult}
\end{center}
\end{figure}
\clearpage
 \section{Summary and perspectives}
In this second paper we have considered many statistical properties of the minimal ABM introduced in \cite{paperoI,paperNP}. We have also considered the generalization of the model to the more realistic case of multiplicative dynamics.\\
The model was introduced to focus on the minimal elements which are able to reproduce the main SF observed in real markets and to discuss their self-organization. These elements are essentially: chartist agents, fundamentalists agents, herding dynamics and price behavior. From the studies we present, we can make some comments about the role of each of these elements. For example, the fundamentalist-chartist competition corresponds essentially to the competition between stabilizing and destabilizing tendencies on the price. We believe that our specific assumptions to describe these tendencies are not crucial for the general behavior and that the results of the model would be robust as soon as these destabilizing and stabilizing tendencies are present in some form. On the other hand the herding effect is absolutely crucial and represents the key element for the dynamics of the Fat Tails and their self-organization in this class of models. Therefore it would be very important to find a way to test the concept of herding in real markets.
The specific price behavior in terms of F and C strategies is important in defining the general properties of the strategies (stability versus instability) but it is not crucial with respect to the specific nature of the SF. \\
We have mostly considered the model within its linear dynamics because it represents a powerful simplification in view of analitical results and interpretation of the data. The more general multiplicative dynamics is shown to be extremely more sensitive to the parameter range and much more difficult to study in all respects. However, the main results for the origin of the SF and the their self-organization, which we have obtained from the linear dynamics, have been also confirmed by the non linear dynamics. The main difference is that the non linear dynamics enhances the deviation form the gaussian behavior and makes the SF more marked. \\
The phenomenon of self-organization interpretated as Self Organized Intermittency (SOI) and Finite Size Effects are fully confirmed also by the non linear dynamics. \\
Our minimal ABM can be easily generalized to consider realistic elements which are important in the real data but that are not considered crucial (in our perspective) to clarify the origin and the self-organization of the SF. One of these elements is the value of the fundamental price $p_f$ which we consider as constant (or zero in the linear model). It is easy to introduce a suitable RW for $p_f(t)$ which permits to understand the dissapearance of the SF and the transition to the gaussian behavior which is observed in real data at long times. \\
It is easy to consider other possible improvements to make the model more realistic. Given the microscopic understanding we have achieved these can be now analysed one by one in a systematic way. For example an element which is considered important \cite{lastbouc,FFFF} and which we intend to study in the future is the problem of the finite liquidity in the market. Finite liquidity induces a higher sensitivity of the price increment (Walras' Law) to the excess demand. A study of a model for the order book would permit to relate the liquidity to the effective number of agents N. If one would know this dependence one could insert it in the coefficient of the Walras' Law, leading to an additional feedback of the dynamics with respect to $N$.\\As this example shows many other specific elements can now be addressed in a systematic way.\\
One point which is left out from our four basic ingredients is the question of the performance of the agents and their persistence in the markets. Our agents can indeed change their startegies on the basis of the past price time series and on the herding phenomenon. On the other hand we limit the strategies to the two broad classes of fundamenlists and chartists. This is a drastic simplification with respects to models which permit a much broader choice of strategies which are selected by genetic or neural algorythms \cite{reviu, hommes, lebaron}. Our model shows, a posteriori, that the simplification in the two broad classes for the two strategies of stabilization and destabilization is enough to achieve a detailed understanding of the SF and their self-organization. On the other hand, the possibility of a broader class of strategies and their choice on the basis of previous performance is certainly an interesting point. The problem, in this respect, is that if the price dynamics is highly  simplified without all the details of an order book, the analysis of the performance can become highly misleading because unrealistic strategies could appear as very profitable. In our opinion, therefore, the question of the strategy selection on the basis of performance is an interesting one but it would require a much higher degree of realism in order to be properly addressed.  

\bibliographystyle{hunsrt}
\bibliography{art}

\appendix
\section{How to define returns}\label{appret}
\begin{figure}[htbp]
\begin{center}
\includegraphics[scale=0.40]{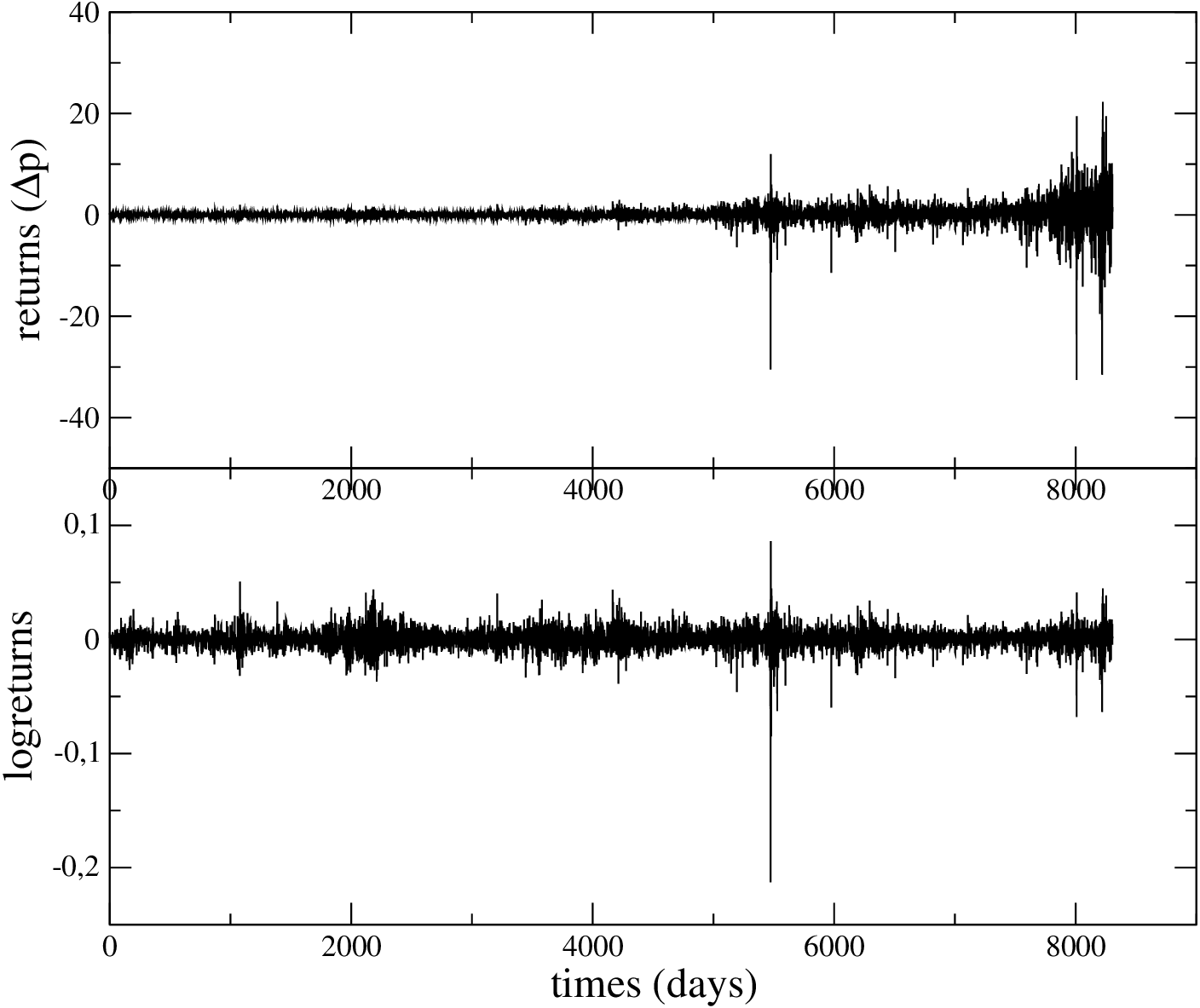}
\caption{Typical daily NYSE index. A visual comparison suggests that the multiplicative hypothesis for prices seems to be better assumption (stationary) rather than the linear one.}
\label{nyse}
\end{center}
\end{figure}

\begin{figure}[!t]
\begin{center}
\includegraphics[scale=0.4]{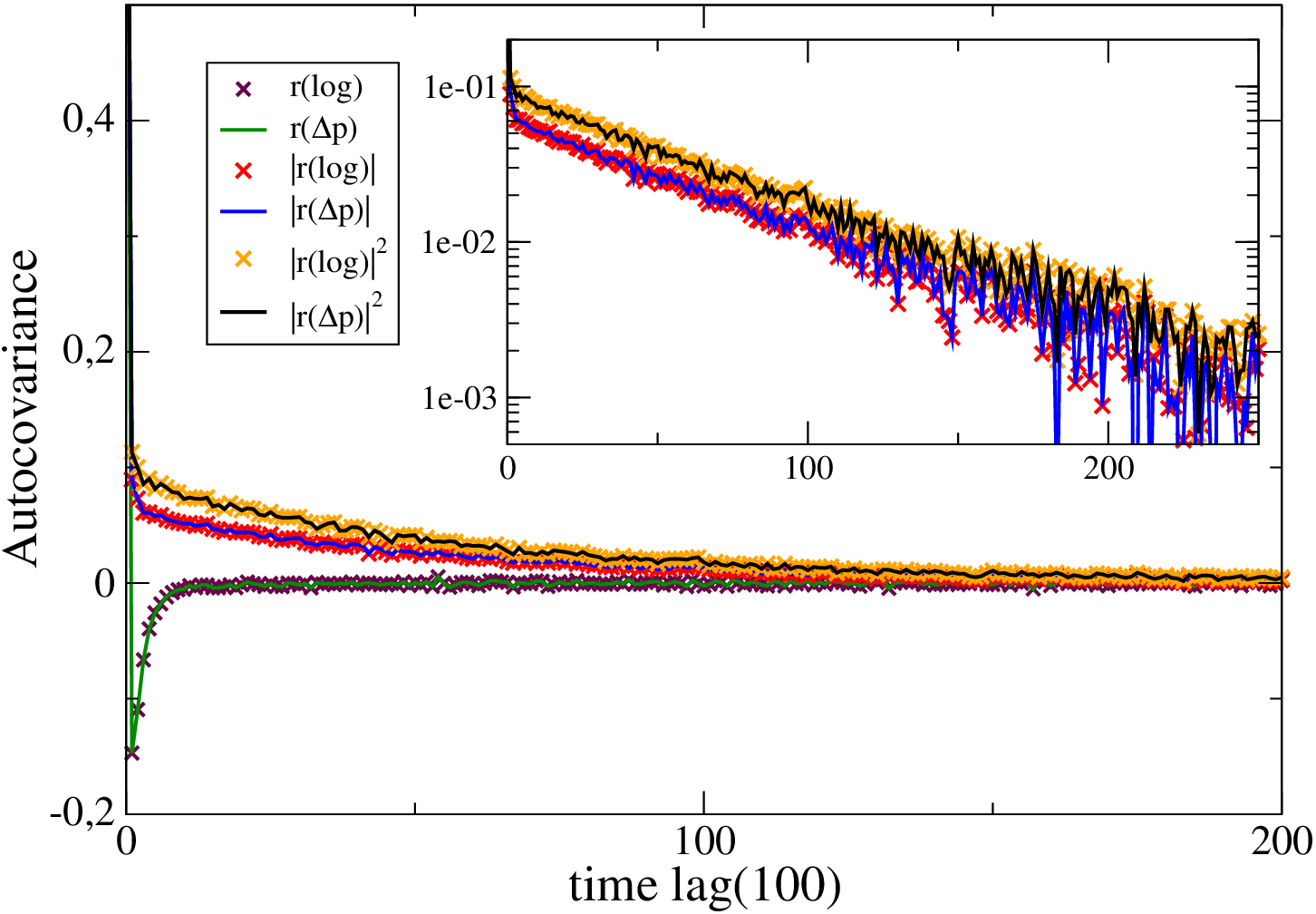}
\caption{The autocovariance function of returns, of absolute returns and of squared returns do not depend on the defintion of the returns in our model because the two functions are nearly equal in both cases. In the insert we highlight that even in semi-logaritmic scale the functions are nearly the same. Here we consider the model with the linear dynamics.}
\label{cfrret}
\end{center}
\end{figure}

In economics returns at a $\Delta$-scale are defined in general as log returns, that is
\begin{equation}
r_{t+\Delta} = \ln\frac{p_{t+\Delta}}{p_t}
\end{equation}
instead in present papers (except for section \ref{moltip}) we have defined returns as
\begin{equation}\label{linret}
r_{t+\Delta} = p_{t+\Delta} - p_t.
\end{equation}
In this appendix we discuss the reason for this choice. 
The log hypothesis is consistent with the idea that it is the logarithm of an asset price to perform a RW rather than the price itself. Morever in such a way the price is defined positive. In Fig. \ref{nyse} we report the daily NYSE returns defined as difference of prices and as log returns. A simple visual inspection confirms the previous statement because only the log returns seem to be reasonably stationary. \\
\\
Anyway we can always write the price at any time as $p_{t+\Delta}=p_t+\Delta p_\Delta $ and, if the time scale $\Delta$ is quite small (from minutes to few hours), we can generally assume that $\Delta p_\Delta \ll p_t$ and consequently:
\begin{equation}\label{Deltap}
r_{t+\Delta} = \ln\bigg(1+\frac{\Delta p_\Delta}{p_t}\bigg)\approx \frac{\Delta p_\Delta}{p_t}\sim \Delta p_\Delta
\end{equation}
because $p_t$ is a slow variable at this time scale (see chapter 5 of \cite{mantegnabook}).
Equation  (\ref{Deltap}) suggests that the way to define returns in our model is irrelevant unless $\Delta p_\Delta$ is very large. For example in Fig. \ref{cfrret} we show the autocovariance function of the returns corresponding to the linear dynamics but considering different definiton of the returns themselves.  The results are extremely similar for the various definitions.\\
\section{Samples and tails}
Let us suppose that the increments are independent and the increment pdf $f(r)$ is symmetric and with zero mean (the assumption of independence is not realistic however it can give some insight). Let us also define the thershold $|r^*|$ as the value above which we define an event as rare. The probability of a rare event is consequently:
\begin{equation}
P(|r|>|r^*|)=2\int_{r^*}^{+\infty}f(r)\,dr:=\beta(r^*)
\end{equation}
The probability $P_r$ of observing at least one rare event among N events is:
\begin{equation}\label{prare}
P_r (N,\beta)= 1 - (1-\beta(r^*))^N\approx 1-e^{-N\beta(r^*)}
\end{equation}
because generally $\beta(r^*)\ll1$. This probabilty is approximately $1$ for $\beta>N^{-1}$ and it is nearly $0$ where $\beta \ll N^{-1}$. So $\beta(r^*)$. fixes the value of $N$ in order to observe rare events. In practice in order to observe rare events $N$ should be larger than $\beta(r^*)^{-1}$. So, if one considers relatively long times correlation the effective size of the sample (with respect to the time interval in question) becomes relatively small in terms of the number of effective steps of a RW.  
\section{Pdf tails}
\begin{figure}[!b]
\begin{center}
\includegraphics[scale=0.4]{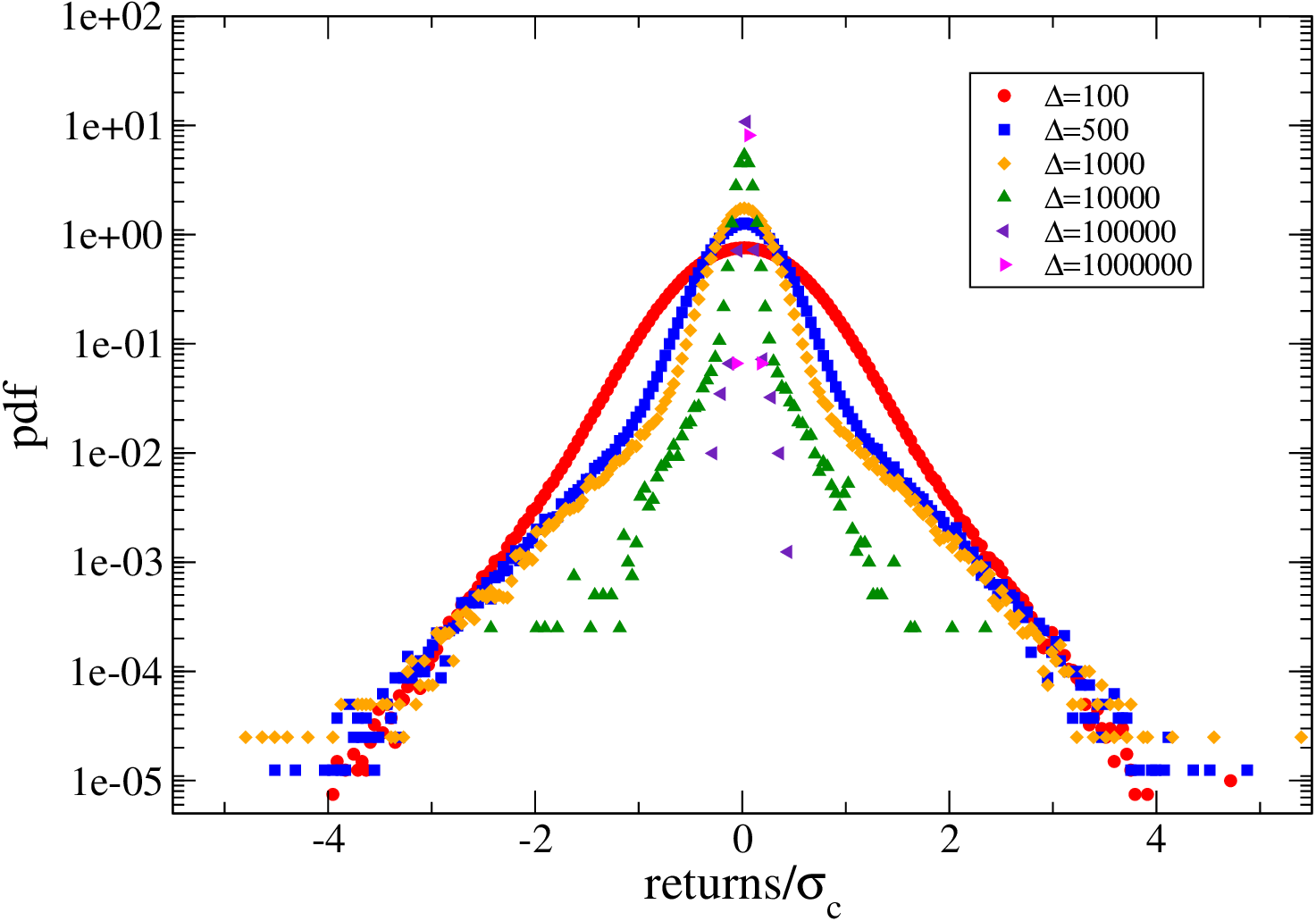}
\caption{Here we show the pdf normalized with the variance of the case $x=1$. The data collapse is not quite satisfactory showing that the scaling behavior of the non gaussian tails is related to the complex fluctuations in the population dynamics and it cannot be interpreted with simple arguments. }
\label{pdfnormchar}
\end{center}
\end{figure}

In Fig. \ref{allpdf2} we have pointed out that the variance of returns does not enterely capture the scaling behavior of the whole return pdf. The motivations may be traced in the origin of the non gaussian tails due to the interaction between a pdf which tends to have constant width and one with increasing width. Fig. \ref{pdfCFN} suggest that tails are somewhat streched by the gaussian with increasing variance (red curves). Conversely the central part, which gives the main contribution to the variance, tends to behave like the fundamentalist gaussian. Therefore, dividing returns with their variance makes that pdfs collapse into a unique curve in the central part. On the other hand, in order to have a collapse of the tails, the variance of the chartist gaussian would appear as a  more natural choice. 
On the other hand Fig. \ref{pdfnormchar} gives only a partial confirmation of this picture. In fact the tails collapse in the same curve only for $\Delta < 1000 - 2000$, we recall that these non collapsing regime (i.e. large value of $\Delta$) corrisponds to the region in which the approximation of section 5 is no more in good quantitative agreement with the simulated results. Therefore it seems that for small value of $\Delta$, the tails grow as quick as the variance of the chartist case but for large $\Delta$ tails are slower than the chartist gaussian but faster than the fundamentalist gaussian. 

\end{document}